\documentclass[aoas,preprint]{imsart}

\usepackage{amsthm,amsmath,amsfonts}
\usepackage[round]{natbib}
\usepackage[english]{babel}
\usepackage{latexsym}
\usepackage{subfigure}
\usepackage{epsfig}
\usepackage{color,soul}
\usepackage{moreverb}
\usepackage{mathtools}
\usepackage{tikz, color}
\usepackage{enumerate,linegoal}
\usepackage{calc}
\usepackage{latexsym}
\usepackage{amsmath}
\usepackage{amsfonts}
\usepackage{amssymb}
\usepackage{bm}
\usepackage{psfrag}
\usepackage{graphicx}
\usepackage{epstopdf}
\usepackage{framed}
\usepackage{booktabs}
\usepackage[flushleft]{threeparttable}

\usepackage{xfrac,array,booktabs}
\usepackage{verbatim}
\usepackage{enumitem}
\usepackage{pdflscape}
\usepackage{rotating}

\DeclareMathOperator*{\argmin}{arg\,min}

\usetikzlibrary{matrix}

\startlocaldefs
\newcommand{\tp}{^{\top}}

\newcommand{\expect}{\operatorname{E}}
\newcommand{\var}{\operatorname{var}}
\newcommand{\cov}{\operatorname{cov}}

\newtheorem{thm}{Theorem}

\newtheorem{corollary}{Corollary}
\endlocaldefs

\begin{document}

\begin{frontmatter}

\title{Mid-quantile regression for discrete responses}
\runtitle{Mid-quantile regression for discrete responses}

\begin{aug}
\author{\fnms{Marco} \snm{Geraci}\corref{}\thanksref{t1}\ead[label=e1]{marco.geraci@uniroma1.it}}
\and
\author{\fnms{Alessio} \snm{Farcomeni}\thanksref{t2}\ead[label=e3]{alessio.farcomeni@uniroma2.it}}

\thankstext{t1}{Corresponding author: Marco Geraci, MEMOTEF Department, School of Economics, Sapienza University of Rome, Via del Castro Laurenziano 9, Rome 00161, Italy. \printead{e1}}

\runauthor{Geraci and Farcomeni}

\affiliation{Sapienza University of Rome and University of South Carolina\thanksmark{t1}, University of Rome `Tor Vergata'\thanksmark{t2}}

\end{aug}

\begin{abstract}
\quad We develop quantile regression methods for discrete responses by extending Parzen's definition of marginal mid-quantiles. As opposed to existing approaches, which are based on either jittering or latent constructs, we use interpolation and define the conditional mid-quantile function as the inverse of the conditional mid-distribution function. We propose a two-step estimator whereby, in the first step, conditional mid-probabilities are obtained nonparametrically and, in the second step, regression coefficients are estimated by solving an implicit equation. When constraining the quantile index to a data-driven admissible range, the second-step estimating equation has a least-squares type, closed-form solution. The proposed estimator is shown to be strongly consistent and asymptotically normal. A simulation study shows that our estimator performs satisfactorily and has an advantage over a competing alternative based on jittering. Our methods can be applied to a large variety of discrete responses, including binary, ordinal, and count variables. We show an application using data on prescription drugs in the United States and discuss two key findings. First, our analysis suggests a possible differential medical treatment that worsens the gender inequality among the most fragile segment of the population. Second, obesity is a strong driver of the number of prescription drugs and is stronger for more frequent medications users. The proposed methods are implemented in the \texttt{R} package \texttt{Qtools}.
\end{abstract}



\end{frontmatter}

\section{Introduction}
\label{sec:1}

In its classical formulation \citep{Koenker1978}, quantile regression (QR) provides a distribution-free approach to the modeling and estimation of quantile treatment effects (QTEs) for \emph{continuous} response variables \citep{Koenker2005}. QR has become a successful analytic method in many fields of science because of its ability to draw inferences about individuals that rank below or above the population conditional mean. The ranking within the conditional distribution of the outcome can be considered as a natural index of individual latent characteristics which cause heterogeneity at the population level \citep{Koenker2001}. The value of estimating QTEs in medical and public health research has been illustrated in several studies \citep{Austin2005,Beyerlein2014,Ding2010,Mayfield2021,Rehkopf2012,Wei2015,Winkelmann2006}.

While most of the progress in QR methods has revolved around continuous responses, relatively less contributions have been made in the discrete case so far. Discrete response variables are ubiquitous in medical research and the literature is arguably dominated by generalized linear models (GLMs) \citep{McCullagh1989}. The reasons are manifold and include convenient interpretation of the regression coefficients, e.g., as (log) odds ratios in logistic regression or (log) rate ratios in Poisson regression, universal availability in statistical software, and the benefits of a well-developed, unifying maximum likelihood theory. However, research has been increasingly directed toward the development of nonparametric (distribution-free) methods to overcome situations in which traditional approaches are unsatisfactory or, more in general, when the goal of the inference transcends the conditional mean of the response.

QR for discrete responses presents some hurdles. Major hindrances include lack of a general theory for handling different types of discreteness, practical estimation challenges, and the troublesome asymptotic behavior of sample quantiles in the presence of ties. Thus, it is not surprising that existing approaches to discrete QR rely on some notion of continuity, either postulated or artificially induced. Early works in the former category date back to the 1950's \citep{Rosenblatt1958}.

Prominent in the econometric literature, maximum score estimation deals with conditional median models of binary \citep{Manski1975,Manski1985,Horowitz1992} and ordered discrete \citep{Lee1992} response variables. More recently, \cite{Kordas2006} extended \citeauthor{Horowitz1992}'s (\citeyear{Horowitz1992}) estimator for binary outcomes to quantiles other than the median. In the maximum score estimation approach, the key assumption is the existence of a continuous latent variable, say $Y^{*}$, which undergoes the working of a threshold mechanism resulting in the observable binary outcome $Y = I(Y^{*} > 0)$, where $I(\cdot)$ denotes the indicator function. The conditional quantiles of the observable outcome are obtained as transformed quantiles of the latent outcome. However, maximum score estimation is computationally expensive as it involves nonconvex loss functions. Jittering is another strategy used for quantile estimation with discrete responses. A (pseudo)continuous variable, say $Z$, is obtained by adding random noise, say $U$, to the observable discrete outcome, i.e. $Z = Y + U$. Estimation then proceeds by applying standard algorithms for convex quantile loss functions (e.g., linear programming) and, successively, by averaging the noise out. This approach, which has been adopted for modeling count \citep{Machado2005} and ordinal \citep{Hong2010} response variables, may lack generality as it requires that adjacent values in the support of $Y$ are equally spaced. Another estimation method for quantiles of count data has been recently proposed \citep{Frumento2021}. An approximation to continuity is introduced in the sense that the response becomes $Z_{E} = Y + \expect(U)$, where $\expect(U) = 0.5$. As explained by the authors, the resulting estimator is asymptotically equivalent whether it is applied to $Z_{E}$ or the jittered response $Z$ as defined above. Efficiency gains of this approach derive from the parametric modeling of a sequence of quantiles, but they are offsetted by the risk of overfitting \citep{Frumento2021} and the inability to model a single quantile.

For the sake of completeness, we briefly mention possible alternatives. Recently, \cite{Chernozhukov2019} proposed inference methods to construct simultaneous confidence bands for quantile and quantile effect functions of possibly discrete random variables. However, their study does not provide any strategy for regression modeling or point estimation. Other proposed methods for estimating non-central summaries of discrete responses include M-quantile regression \citep{Chambers2014,Chambers2016}. Unfortunately, conditional M-quantiles suffer from lack of interpretability as their relationship with conditional quantiles is obscure. They also depend on global properties of the parent distribution, despite they attempt to describe a local property of such distribution \citep{Koenker2013}. Another class of models that may be used to estimate conditional quantiles can be collectively referred to as distributional regression. Whether distributional regression is used within a non-, semi- or fully parametric framework, the common goal is to model the conditional distribution function (CDF) as flexibly as possible by means of covariate-dependent distributional parameters. For example, covariates may enter in the model via parameters related to the location, scale and shape of the distribution \citep{Stasi2018}. While the conditional quantile function can be obtained by inverting the conditional CDF (though in general this is not guaranteed to yield an analytical expression), the former will typically depend on the regression parameters in a complicated (often implicit and nonlinear) fashion. That is, distributional regression parameters may not give a simple and straightforward summary of the QTE, in contrast to the parameters of QR models \citep{Koenker2005}. Finally, in the hope of remedying a misconception that has crept into the literature of quantile regression, we note that the method proposed by \cite{Bottai2009}, despite being called `logistic quantile regression', is applied to \emph{continuous} responses only and has nothing to do with conditional quantiles of binary responses.

In this paper, we build on mid-quantiles \citep{Parzen1993} to introduce an alternative estimation approach for conditional quantiles of discrete responses. Sample mid-quantiles, which are based on essentially the same idea of the mid-$p$-value \citep{Lancaster1961}, offer a unifying theory for quantile estimation with continuous or discrete variables and are well-behaved asymptotically \citep{Ma2011}. In our approach, we develop a two-step estimator that can be applied to a large variety of discrete responses, including binary, ordinal, and count variables, and is shown to have good theoretical properties. In a simulation study, we gather empirical evidence that conditional mid-quantile estimation is more efficient than jittering. However, this evidence is contextual and may not be generalizable.

The rest of the paper is organized as follows. In the next section we discuss modeling, estimation, and theoretical properties of conditional mid-quantile estimators, with technical details on inference given in Appendix~\ref{sec:A}. In Section~\ref{sec:3}, we report the results of a simulation study to assess bias and efficiency of the proposed estimator, as well as confidence interval coverage. We also illustrate an application to data on prescription drugs use in the United States. We conclude with final remarks in Section \ref{sec:4}.

\section{Methods}
\label{sec:2}
\subsection{Marginal mid-quantiles}
\label{sec:2.1}
Let $Y$ be a discrete random variable with probability mass function $m_{Y}(y) = \Pr(Y = y)$ and cumulative distribution function (CDF) $F_{Y}(y)=\sum_{u \leq y} m_Y(u)$. The $p$th quantile of $Y$, denoted by $\xi_{p}$, is defined as $\xi_{p} \equiv \inf\{y \in \mathbb{R}: F_{Y}(y)\geq p\}$ for any $0 < p < 1$. We may define the quantile function (QF) of $Y$, $Q_{Y}(p)$, as the generalized inverse of the CDF of Y, that is $Q_{Y}(p) \equiv F_{Y}^{-1}(p)$. In the discrete case, the CDF is not injective, thus a discrete QF is not the standard inverse of the CDF. Now let $Y_{1}, Y_{2}, \ldots, Y_{n}$ be an independent sample of size $n$ from the population $F_{Y}$. The sample CDF is defined as $\hat{F}_{Y}(y) = n^{-1}\sum_{i=1}^{n} I\left(Y_{i} \leq y\right)$, $y \in \mathbb{R}$, while the (ordinary) sample QF, defined as the inverse of the sample CDF \citep[see, for example,][for a detailed overview of alternative sample quantiles]{Hyndman1996}. The sample QF, too, is discrete.

In general, sample quantiles as defined above may not be consistent for the population quantiles when the underlying distribution is discrete \citep{Jentsch2016}. Additionally, the sample median $\hat{Q}_{Y}(0.5) = Y_{(\lceil n/2 \rceil)}$ lacks asymptotic normality if $Y$ is discrete \citep{Genton2006}. Throughout this article, we use $\lceil x \rceil$ ($\lfloor x \rfloor$) to denote the smallest (largest) integer that is larger (smaller) than or equal to $x$.

We now introduce the mid-cumulative distribution function (mid-CDF) \citep{Parzen1993,Parzen2004}, a modification of the standard CDF that plays an important role in discrete modeling and in samples with ties. For a random variable $Y$ with CDF $F_{Y}(y)$, the function
\begin{equation}\label{eqn:1}
G_{Y}(y) \equiv \Pr(Y \leq y) - 0.5\cdot \Pr(Y = y)
\end{equation}
is called mid-distribution function (mid-CDF). Since $Y$ is discrete, $G_{Y}(y)$ is a step function (a downward-shifted version of $F_{Y}(y)$). Note that, if $Y$ were instead continuous, then $G_{Y}(y)$ would reduce to $F_{Y}(y)$ since, in that case, $\Pr(Y = y)= 0$.

\begin{figure}[t!]
\centering
\includegraphics[scale = 0.4]{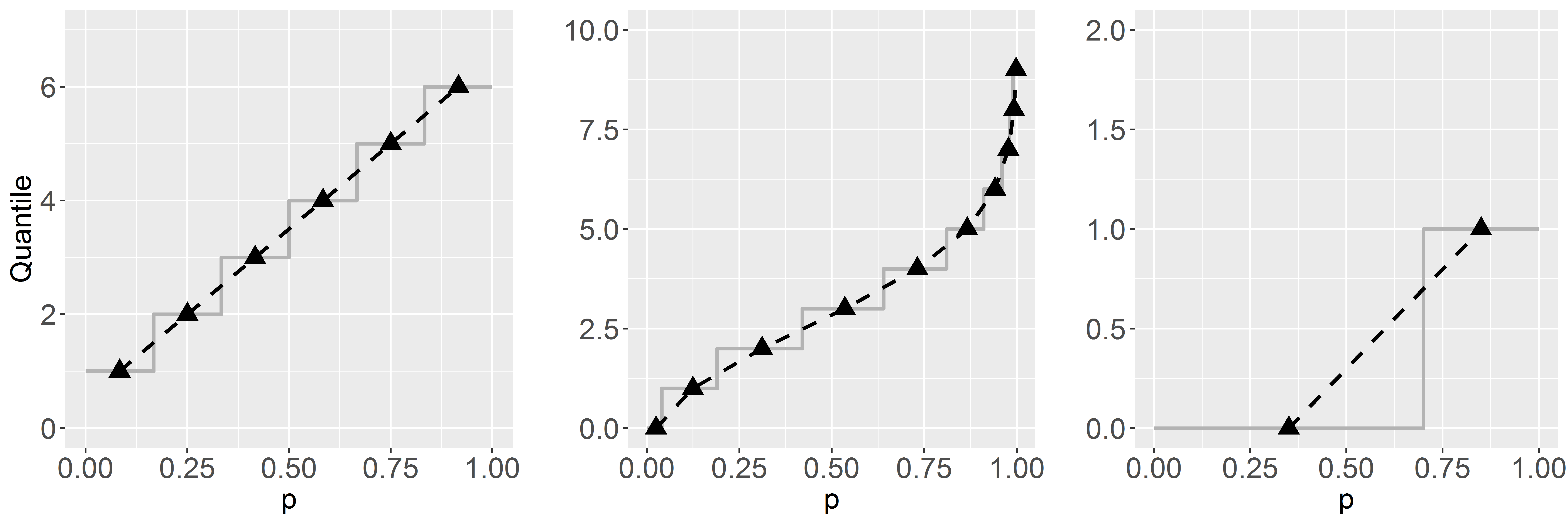}
\caption{True quantile function (grey solid line) and mid-quantiles (black filled triangles). Left: discrete uniform on (1,6). Center: Poisson with mean 3. Right: Bernoulli with probability 0.3. \label{fig:1}}
\end{figure}

Further, let $\mathcal{S}_{Y} = \{y_{1}, \ldots, y_{s}\}$, {with $y_{j} < y_{j + 1}$ for all $j = 1, \ldots, s - 1$}, be the set of $s$ distinct values in the population that the discrete random variable $Y$ can take on, with corresponding probabilities $p_{1},\ldots,p_{s}$. We also define the mid-probabilities $\pi_{1} = p_{1}/2$ and $\pi_{j} = G(y_{j}) = \sum_{u = 1}^{j - 1}p_{u} + p_{j}/2$, for $j = 2,\ldots, s$. The following function
\begin{equation}\label{eqn:2}
H_{Y}(p) =
\begin{cases}
    y_{1}, & \text{if $p < \pi_{1}$,}\\
    y_{j}, & \text{if $p = \pi_{j}$, $j = 1,\ldots,s$,}\\
    (1- \gamma) y_{j} + \gamma y_{j + 1}, & \text{if $p = (1- \gamma) \pi_{j} + \gamma \pi_{j + 1}$,}\\
    & \text{$0 < \gamma < 1$, $j = 1,\ldots,s - 1$,}\\
    y_{s}, & \text{if $p > \pi_{s}$,}\\
\end{cases}
\end{equation}
is called mid-quantile function (mid-QF) \citep{Ma2011}. If $s = \infty$, then the last category is suppressed. Examples of $H_{Y}(p)$ when $Y$ is discrete uniform, count, or binary are given in Figure~\ref{fig:1}. The mid-QF is piecewise linear and connects the points $\left(\pi_{j},y_{j}\right)$ (dashed lines in Figure~\ref{fig:1}). One can verify that $H_{Y}\left\{G_{Y}(y_{j})\right\} = y_{j}$, $j = 1,\ldots,s$. In general, mid-quantiles cannot be obtained by inverting $G_{Y}(y)$ at points $y \not\in\mathcal{S}_{Y}$. However, we can define $G_{Y}^{c}(y)$, the continuous version of $G_{Y}(y)$ \citep{Parzen2004}, as the piecewise linear function that connects the values $G^{c}_{Y}(y_{j}) = \pi_{j}$, $j = 1, \ldots, s$, and satisfies $G_{Y}^{c}(y) \equiv H_{Y}^{-1}(y)$, for all $y \in \mathbb{R}$. Related to this, we have the equivariance property $h^{-1}\left\{H_{h(Y)}(\pi_{j})\right\} = H_{Y}(\pi_{j})$, $j = 1,\ldots,s$, for a monotone transformation $h$. Equivariance of $H_{Y}(p)$ no longer applies if $H_{Y}(\pi_{j})< p <H_{Y}(\pi_{j+1})$ unless $h$ is linear.

The sample mid-CDF corresponding to \eqref{eqn:1} is $\hat{G}_{Y}(y) = \hat{F}_{Y}(y) - 0.5\cdot \hat{m}_{Y}(y)$, where $\hat{m}_{Y}(y) = n^{-1}\sum_{i=1}^{n} I\left(Y_{i} = y\right)$ is the sample relative frequency of $y$. To define the sample mid-quantiles, we need to introduce some more notation. Let $z_{j}$, $j = 1, \ldots, k$, be $k$ distinct values that occur in the sample, {with $z_{j} < z_{j + 1}$ for all $j = 1, \ldots, k - 1$}, and let $\hat{p}_{j} = \hat{m}_{Y}(z_{j})$, $j = 1, \ldots, k$, be the corresponding relative frequencies. Then $\hat{H}_{Y}(p) = (1-\gamma) z_{j} + \gamma z_{j + 1}$, where {the index $j = 1,\ldots,k-1$} is such that $p = (1- \gamma) \hat{G}_{Y}(z_{j}) + \gamma \hat{G}_{Y}(z_{j + 1})$, $0 \leq \gamma \leq 1$, and $\hat{G}_{Y}(z_{j}) = \hat{F}_{Y}(z_{j}) - 0.5\cdot \hat{p}_{j}$. Moreover, $\hat{H}_{Y}(p) = z_{1}$ if $p < \hat{G}_{Y}(z_{1})$ and $\hat{H}_{Y}(p) = z_{k}$ if $p > \hat{G}_{Y}(z_{k})$. A natural estimator of $G^{c}_{Y}(y)$ is then $\hat{G}^{c}_{Y}(y) = (1-\gamma) \hat{\pi}_{j} + \gamma \hat{\pi}_{j + 1}$ for $z_{j} \leq y \leq z_{j+1}$.

In samples with ties, $\hat{H}_{Y}(p)$ is the piecewise linear function connecting the values $\hat{H}_{Y}\{\hat{G}_{Y}(z_{j})\} = \hat{H}_{Y}\{\hat{G}^{c}_{Y}(z_{j})\} = z_{j}$. It has been showed that if the underlying distribution $F_{Y}$ is absolutely continuous, then the sample mid-quantiles have the same asymptotic properties as the ordinary sample quantiles \citep{Ma2011}. More importantly, if $F_{Y}$ is discrete, then the sample mid-quantiles are consistent estimators of the population mid-quantiles and their sampling distribution is normal \citep{Ma2011}.

As far as interpretation goes, mid-quantiles can be viewed as fractional order statistics \citep{Stigler1977,Genton2006}. A mid-quantile is a quantile of $Y$ when $p = \pi_{j}$, $j = 1, \ldots, k$. Otherwise, when $p \neq \pi_{j}$, its interpretation involves an underlying form of continuity that captures the smooth progression from one quantile to the next. It is in this spirit that \cite{Wang2011} proposed smooth quantiles for discrete distributions based on fractional order statistics. Compared with ordinary sample quantiles, it can be argued that mid-quantiles offer a sensible approach to quantifying the differences between discrete distributions. For example, consider two binary samples with different proportions of successes, $\{0,0,0,0,1\}$ and $\{0,0,0,1,1\}$ \citep{Ma2011}. The sample median is $0$ in both cases; the sample mid-median corresponds to the proportion of $1$'s, that is, $1/5$ in the former sample, but $2/5$ in the latter. In this example, the mid-median helps discriminate between the two samples in an intuitive way (as formally shown in Section~\ref{sec:2.2}, our proposed conditional mid-median inherits the same properties).

Another advantage of mid-quantiles is that they can be easily relabeled to obtain the (ordinary) quantiles. Suppose $\hat{H}_{Y}(p) = z_{j}$ and the goal is to recover the sample quantile $\hat{\xi}_{p}$. First, the mid-probabilities are obtained from the inversion $\hat{H}^{-1}_{Y}\left(z_{j}\right) = \hat{G}_{Y}(z_{j})$, while the sample CDF is calculated recursively from $\tilde{F}(z_{j}) = 2\hat{G}_{Y}(z_{j}) - \tilde{F}_{Y}(z_{j-1})$, $j = 1, \ldots,k$, with the convention that $\tilde{F}(z_{0}) = 0$. This leads to $\hat{\xi}_{p} = z_{j}$ for $\tilde{F}(z_{j-1}) < p \leq \tilde{F}(z_{j})$, $j = 1, \ldots,k$. Otherwise, if $\hat{H}_{Y}(p) = z \neq z_{j}$, then one takes $\hat{H}^{-1}_{Y}\left(z_{j^{*}}\right)$ where $z_{j^{*}}$ is the largest of the $z_{j}$'s that satisfy $z > z_{j}$. This is an important property that sets mid-quantiles apart from so-called `quantile-like' alternatives like expectiles. The latter have a non-trivial relationship with quantiles \citep{Jones1994} and require non-trivial relabeling procedures that may lead to poor approximations on the tails \citep{Koenker2013}. M-quantiles, which represent a generalization of expectiles, suffer from a similar limitation.

\subsection{Conditional mid-quantiles}
\label{sec:2.2}

Analogously to \eqref{eqn:1}, we define the \emph{conditional} mid-CDF as
\begin{equation}\label{eqn:3}
G_{Y|X}(y|x) \equiv F_{Y|X}(y|x) - 0.5\cdot m_{Y|X}(y|x),
\end{equation}
where $Y$ is a random variable with image $\mathcal{S}_{Y} \subset \mathbb{R}$ and $X$ is a $q$-dimensional vector of covariates, which may include a constant equal to one for the intercept, $F_{Y|X}(y|x) = \Pr(Y \leq y|X = x)$, and $m_{Y|X}(y|x) = \Pr(Y = y|X = x)$. Although the definition of conditional mid-CDF applies to both continuous and discrete response variables (as in the case of the marginal mid-CDF \eqref{eqn:1}), we assume that $\mathcal{S}_{Y} = \{y_{1}, \ldots, y_{s}\}$ is a finite or countably infinite ($s = \infty$) subset of $\mathbb{R}$. In particular, $Y$ can be binary, ordinal, or count, with positive or negative values, not necessarily equally spaced. Values need not be integers either. However, we do exclude nominal variables with more than two categories from the application of \eqref{eqn:3} as they do not have a natural ordering.

We use the conditional mid-CDF \eqref{eqn:3} as the springboard for defining conditional mid-quantiles. Let $\pi_{j} = G_{Y|X}(y_{j}|x)$ (for the sake of simplicity, we have suppressed the dependence on $x$ from $\pi_{j}$'s notation). The \emph{conditional} mid-QF $H_{Y|X}(p)$ is defined as the piecewise linear connecting the values $G^{-1}_{Y|X}(\pi_{j}|x)$, $j = 1, \ldots, s$, for given $x$. We assume a quantile-specific model that is linear on the scale of $h$, i.e.,
\begin{equation}\label{eqn:4}
H_{h(Y)|X}(p) = x\tp\beta(p),
\end{equation}
where $h$ is a known monotone and differentiable `link' function, and $\beta(p)$ is a vector of $q$ unknown regression coefficients for a given $p \in (0,1)$. In our approach, $h$ may simply be the identity or a linear transformation, the logarithmic function---which is typically used in the modeling of counts \citep{Machado2005}, the logistic function, or belong to a family of flexible transformation models \citep{Chamberlain1994,Mu2007,Yin2008,Geraci2015}. These often involve the Box-Cox \citep{Box1964} or Aranda-Ordaz \citep{AO1981} families. As in the marginal case, conditional mid-quantiles cannot be obtained by inverting $G_{Y|X}(y|x)$ at points $y \not\in\mathcal{S}_{Y}$. Again, we define the piecewise linear function $G^{c}_{Y|X}(y|x)$ that connects the values $G^{c}_{Y|X}(y_{j}|x) = \pi_{j}$, $j = 1, \ldots, s$, and satisfies $G_{Y|X}^{c}(y|x) \equiv H_{Y|X}^{-1}(y|x)$, for all $y \in \mathbb{R}$.

\begin{figure}[t!]
\centering
\includegraphics[scale = 0.5]{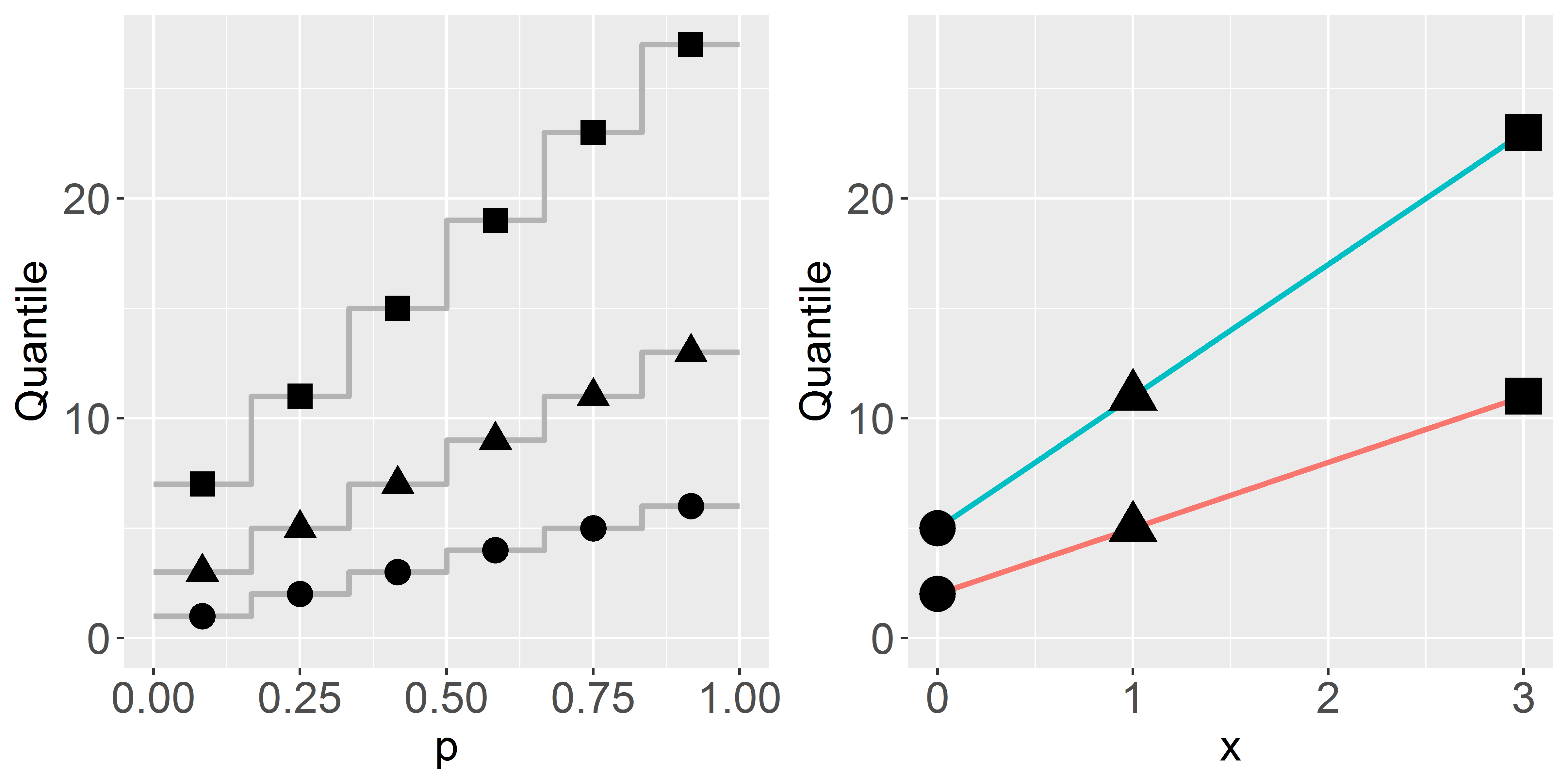}
\caption{True quantile function (grey solid line) and mid-quantiles (black filled circles, triangles and squares). Left: quantiles as a function of $p$ by values of $x \in \{0,1,3\}$. Right: quantiles as a function of $x$ by values of $p \in \{0.25,0.75\}$. \label{fig:2}}
\end{figure}

The QTE interpretation of the $j$th coefficient $\beta_{j}(p)$ is immediate if $x_{j}$ is discrete. Otherwise, it can be defined in terms of the partial derivative of $H$. For example, if $h$ is linear and there are no additional terms that depend on $x_{j}$ (e.g., interactions, quadratic terms), then the QTE associated with $x_{j}$ is given by
$$
\frac{\partial H_{Y|X}(p)}{\partial x_{j}} = \beta_{j}(p).
$$
In Figure~\ref{fig:2}, we provide a simple example using the heteroscedastic model $Y = \lfloor x \rfloor + \lfloor x + 1\rfloor\epsilon$, for $x \geq 0$, where $\epsilon$ is discrete uniform between 1 and 6 (die rolling). As a function of $p$, for fixed $x$, mid-quantiles follow the uniform model we already seen in Figure~\ref{fig:1}. As a function of $x$, for fixed $p$, we obtain our proposed mid-quantile function~\eqref{eqn:4}. In particular, the first and third mid-quartiles have equations $H(0.25) = 2 + 3x$ and $H(0.75) = 5 + 6x$, respectively, which entail a QTE equal to 3 and 6, respectively.

Our definition of conditional mid-quantiles is general as it applies to any type of discrete response variable that can be ordered. An interesting special case is when $Y$ is binary, thus $\mathcal{S}_{Y} = \{0,1\}$. According to \eqref{eqn:3}, the mid-CDF is then given by $G_{Y|X}(y|x) = (y+1-\mu(x))/2$, where $\mu(x) = \Pr(Y = 1|X = x)$. Therefore, $\pi_{1} = G_{Y|X}(0|x) = 0.5-0.5\mu(x)$ and $\pi_{2} = G_{Y|X}(1|x) = 1-0.5\mu(x)$. By definition, mid-quantiles are equal to 0 if $p < \pi_{1}$ and to 1 if $p > \pi_{2}$. Otherwise, $H_{Y|X}(p) = (1-\gamma)y_{1} + \gamma y_{2} = \gamma$ for $p = (1- \gamma) \pi_{1} + \gamma \pi_{2}$. From the latter expression we get $\gamma = (p - \pi_{1})/(\pi_{2} - \pi_{1}) = 2p - 1+\mu(x)$. In summary, we obtain the following conditional mid-quantile function
\begin{equation*}
H_{Y|X}(p) = \begin{cases}
0 & \mbox{if $p < \pi_{1}$,}\\
2p - 1 + \mu(x) & \mbox{if $\pi_{1} \leq p \leq \pi_{2}$,}\\
1 & \mbox{if $p > \pi_{2}$,}\\
\end{cases}
\end{equation*}
Therefore the conditional mid-median is exactly equal to $\mu(x)$, while all the other conditional mid-quantiles are shifted by $2p - 1$.

\subsection{Estimation}
\label{sec:2.3}
Consider a sample $(X_{i},Y_{i})$, $i = 1, \ldots,n$, with corresponding observations $(x_{i},y_{i})$. Also, let $z_{j}$, $j = 1, \ldots, k$, be the $j$th distinct observation of $Y$ that occurs in the sample, with $z_{j} < z_{j + 1}$ for all $j = 1, \ldots, k - 1$.

Estimation of model \eqref{eqn:4} proceeds in two steps. In the first step, we estimate the mid-CDF. Let
\begin{equation}\label{eqn:5}
\hat{G}_{Y|X}(y|x) \equiv \hat{F}_{Y|X}(y|x) - 0.5\cdot \hat{m}_{Y|X}(y|x).
\end{equation}
denote the sample equivalent of \eqref{eqn:3}. The estimation of $F_{Y|X}$ plays a key role in our approach. In our formulation, we require an estimator that can be applied to discrete responses and that admits continuous and discrete covariates (or a mix thereof). In line with the nonparametric flavor of our modeling strategy, we considered the conditional CDF estimator proposed by \cite{Li2008}. This takes the form
\begin{equation}\label{eqn:6}
\hat{F}_{Y|X}(y|x) = \frac{n^{-1}\sum_{i=1}^{n} I(Y_{i} \leq y)K_{\lambda}(X_{i},x)}{\hat{\delta}_{X}(x)},
\end{equation}
where $K_{\lambda}$ is the (product) kernel with bandwidth vector $\lambda$ and $\hat{\delta}_{X}(x)$ is the kernel estimator of the marginal density of $X$. The implementation of \eqref{eqn:6} involves choosing kernel types, as well as determining a number of tuning and estimation parameters. In our simulation study (Section~\ref{sec:3}), we adopted the default settings of the \texttt{np} package \citep{Hayfield2008}, which include, for example, least-squares cross-validation for bandwidth selection \citep{Li2013}. Since we obtained satisfactory empirical results using those settings, we will not dwell on this matter but instead refer the reader to the relevant literature for technical and implementation details \citep{Wang1981, Li2007, Li2008, Li2013, Hayfield2008}.

By applying \eqref{eqn:6} to the sample observations, we obtain $\hat{F}_{Y|X}(z_{j}|x)$, $j = 1, \ldots, k$, and $\hat{m}_{Y|X}(z_{j}|x) = \hat{F}_{Y|X}(z_{j}|x) - \hat{F}_{Y|X}(z_{j-1}|x)$, which we plug into \eqref{eqn:5} to obtain $\hat{G}_{Y|X}(z_{j}|x)$. Here, we set $\hat{m}_{Y|X}(z_{1}|x) = \hat{F}_{Y|X}(z_{1}|x)$, hence $\hat{G}_{Y|X}(z_{1}|x) = 0.5\cdot \hat{F}_{Y|X}(z_{1}|x)$.

We note that, unfortunately, nonparametric estimation of $F_{Y|X}$ entails a loss of performance when the dimension of $X$ is large, the design is sparse, or both. In these cases a semiparametric approach may be preferred. A natural choice is to obtain $\hat{F}_{Y|X}(z_{j}|x)$ as the estimate of the binomial probability $\Pr\left\{I(Y \leq z_{j})|x\right\}$, for $j = 1, \ldots, k$. This idea was indeed considered by some authors \citep{Foresi1995,Peracchi2002} to address the curse of dimensionality of nonparametric estimators and, while originally a logit estimator was proposed, in principle any other link function can be employed. Since applying the binomial estimator is tantamount to fitting $k$ separate binomial regressions, one for each value $z_{j}$, one must verify whether the estimates $\hat{F}_{Y|X}(z_{j}|x)$ are monotone. Monotonicity can be imposed {\it a priori} \citep{Peracchi2002} or {\it a posteriori} \citep{Chernozhukov2010}.

Now we move to the second step of model estimation. For a given $x$, define $\hat{G}_{Y|X}^{c}\left(y|x\right)$ as the function interpolating the points $(z_{j}, \hat{G}_{Y|X}(z_{j}|x))$, where the ordinates have been obtained in the first step. The function $\hat{G}_{Y|X}^{c}$ estimates $G_{Y|X}^{c}$ and is the conditional extension of $\hat{G}_{Y}^{c}$ introduced in Section~\ref{sec:2.1}. The goal is to estimate $\beta(p)$ in \eqref{eqn:4} by solving the implicit equation $p = \hat{G}_{Y|X}^{c}\left(\eta(p)|x\right)$, where $\eta(p) = h^{-1}\{x\tp\beta(p)\}$. Our objective function and estimator are thus given by
\begin{equation}\label{eqn:7}
\psi_{n}(\beta; p) = n^{-1}\sum_{i = 1}^{n} \left\{p - \hat{G}_{Y|X}^{c}\left(\eta_{i}|x_{i}\right)\right\}^{2},
\end{equation}
where $\eta_{i} = h^{-1}\{x_{i}\tp\beta\}$, and
\begin{equation}\label{eqn:8}
\hat{\beta}(p) = \argmin_{\beta \in \mathbb{R}^{q}} \psi_{n}(\beta; p),
\end{equation}
respectively. This estimation approach, which is an extension of the `inverse-CDF' technique for marginal quantiles, has been recently proposed by \cite{deBacker} for fitting censored quantile regression models. They, too, approach nonparametrically the estimation of the CDF (though in the continuous domain) via a double-kernel estimator akin to that of \cite{Li2008}. In summary, we echo \citeauthor{deBacker}'s (\citeyear{deBacker}) advocacy of the squared distance for its computational advantages and of the double-kernel estimation for its asymptotic and finite sample efficiency gains.

We can make \eqref{eqn:8} explicit by using the following linear interpolating function
\begin{eqnarray*}
\hat{G}_{Y|X}^{c}\left(\eta_{i}|x_{i}\right) = b_{j_{i}}(\eta_{i} - z_{j_{i}}) + \hat{\pi}_{j_{i}} & & \qquad \text{$z_{j_{i}} \leq \eta_{i} \leq z_{j_{i}+1}$,}
\end{eqnarray*}
where $b_{j_{i}} = \dfrac{\hat{\pi}_{j_{i}+1} - \hat{\pi}_{j_{i}}}{z_{j_{i}+1} - z_{j_{i}}}$ and $\hat{\pi}_{j_{i}} = \hat{G}_{Y|X}(z_{j_{i}}|x_{i})$. The index $j_{i} = 1,\ldots,k-1$ identifies, for a given  $i = 1, \ldots, n$, the value $z_{j_{i}}$ among the $z$'s such that $\hat{G}_{Y|X}(z_{j_{i}}|x_{i}) \leq p \leq \hat{G}_{Y|X}(z_{j_{i}+1}|x_{i})$. If we restrict $p \in \mathcal{I}$, where $\mathcal{I} = \left[\max_{i}\hat{G}_{Y|X}(z_{1}|x_{i}),\right.$ $\left.\allowbreak \min_{i}\hat{G}_{Y|X}(z_{k}|x_{i})\right]$, then we find that our estimator, conditionally on $\hat{\pi} = (\hat{\pi}_{1_{1}}, \ldots, \hat{\pi}_{k_{1}}, \ldots, \hat{\pi}_{1_{n}}, \ldots, \hat{\pi}_{k_{n}})\tp$, has the form
\begin{equation}\label{eqn:9}
\hat{\beta}(p; \hat{\pi}) = (\mathbf{X}\tp \mathbf{X})^{-1}\mathbf{X}\tp u,
\end{equation}
where $\mathbf{X}$ is a $n \times q$ matrix with $i$th row $x_{i}$ and $u$ is a $n \times 1$ vector with $i$th element $h\left(\frac{p - \hat{\pi}_{j_{i}}}{b_{j_{i}}} + z_{j_{i}}\right)$. It is straightforward to verify that \eqref{eqn:9} is a minimizer by plugging it into \eqref{eqn:7}. The closed-form of \eqref{eqn:9} is, clearly, computationally convenient. Of course, one can still obtain an estimate of $\beta(p)$ using \eqref{eqn:9}, regardless of whether $p$ is within the interval $\mathcal{I}$. However, when $p \not\in \mathcal{I}$, all the elements of $u$ such that $\hat{\pi}_{1_{i}} > p$ or $\hat{\pi}_{k_{i}} < p$ are `censored', that is, constrained to be $h(z_{1})$ or $h(z_{k})$. As a result, the linear predictor of the mid-quantile model will flatten out if $p$ approaches 0 or 1, with `slopes' tending to 0 and the intercept tending to the sample minimum or maximum, at a speed that depends on the censoring rate. This is a reasonable behavior since it reflects all we can say about the relationship between $Y$ and $X$ when we have little or no information. Note also that, in principle, one can consider other general minimizers of \eqref{eqn:7} should the optimization problem become more complex (e.g., because of the addition of nonlinear constraints or penalties). This is a topic for future research.

We can derive the variance-covariance of $\hat{\beta}(p)$ via the total variance law \citep{Mood1974} as follows:
\begin{equation}\label{eqn:10}
\var\left(\hat{\beta}(p)\right) = \expect_{\hat{\pi}}\left\{\var_{\hat{\beta}}(\hat{\beta}|\hat{\pi})\right\} + \var_{\hat{\pi}}\left\{\expect_{\hat{\beta}}(\hat{\beta}|\hat{\pi})\right\}.
\end{equation}
We estimate the first term in the right-hand side of~\eqref{eqn:10} using a Huber-White variance-covariance estimator, which is given by $(\mathbf{X}\tp \mathbf{X})^{-1}\mathbf{X}\tp \mathbf{D} \mathbf{X} (\mathbf{X}\tp \mathbf{X})^{-1}$, where $\mathbf{D} = \mathrm{diag}(\hat{e}^{2}_{1}, \ldots, \hat{e}^{2}_{n})$ and $\hat{e}_{i} = h(y_{i}) - x_{i}\tp \hat{\beta}(p; \hat{\pi})$, $i = 1,\ldots,n$. To obtain an estimate of the second term, we note that, by the delta method,
\begin{equation}
\label{eqn:11}
\var_{\hat{\pi}}\left\{\expect_{\hat{\beta}}(\hat{\beta}|\hat{\pi})\right\} \approx \nabla_{\hat{\pi}} \hat{\beta}(p; \hat{\pi})\tp \var({\hat{\pi}}) \nabla_{\hat{\pi}}\hat{\beta}(p; \hat{\pi}).
\end{equation}
The validity of \eqref{eqn:11} relies on regularity conditions on moments \citep{Oehlert1992}, which in our case are guaranteed by the asymptotic normality established by Theorem~\ref{theo:2}. The expression for $\var({\hat{\pi}})$ depends on the variance of the estimator $\hat{F}_{Y|X}$, which we discuss further below. We omit the tedious algebra for the Jacobian $\nabla_{\hat{\pi}} \hat{\beta}(p; \hat{\pi})$, which can be easily obtained via numerical differentiation. Also, note that the latter is carried out efficiently since the Jacobian is sparse, with sparsity no less than $1 - \frac{2n}{(nk)^2}$. This follows from the fact that the $i$th partial derivatives of $\hat{\beta}(p; \hat{\pi})$ with respect to elements of $\hat{\pi}$ with indices other than $j_{i}$ and $j_{i +1}$ are null (hence, there are at most $2n$ non-zero partial derivatives). The diagonal elements of the $nk \times nk$ matrix $\var({\hat{\pi}})$ are given by
\begin{align*}
\var\left(\hat{\pi}_{ij}\right) = & \, \frac{1}{4}\var\left\{\hat{F}_{Y|X}(z_{j-1}|x_{i})\right\} + \frac{1}{4}\var\left\{\hat{F}_{Y|X}(z_{j}|x_{i})\right\}\\
 & + \frac{1}{2} \cov \left\{\hat{F}_{Y|X}(z_{j-1}|x_{i}),\hat{F}_{Y|X}(z_{j}|x_{i})\right\},
\end{align*}
$i = 1,\ldots,n$, $j = 2, \ldots, k$, and $\var\left(\hat{\pi}_{i1}\right) = \frac{1}{4}\var\left\{\hat{F}_{Y|X}(z_{1}|x_{i})\right\}$. In the expression above, we can neglect the covariance between $\hat{F}_{Y|X}(z_{j}|x_{i})$ and $\hat{F}_{Y|X}(z_{j'}|x_{i})$, $j \neq j'$, as this is asymptotically zero as shown in the proof of Theorem~\ref{theo:2} (Section~\ref{sec:2.5}). This means that the off-diagonal elements of $\var({\hat{\pi}})$ are also asymptotically negligible. Finally, the expression for the variance of $\hat{F}_{Y|X}$ is given elsewhere \citep{Li2008}.

We conclude this section by noting the we would arrive at the estimator \eqref{eqn:9} also by starting from the conditional extension of the marginal mid-quantiles in \eqref{eqn:2}, say
\begin{equation*}
H_{Y|X = x_{i}}(p) =
\begin{cases}
y_{j_{1}}, & \text{if $p < \pi_{j_{1}}$,}\\
y_{j_{i}}, & \text{if $p = \pi_{j_{i}}$, $j_{i} = 1,\ldots,k$,}\\
(1- \gamma_{i}) y_{j_{i}} + \gamma_{i} y_{j_{i} + 1}, & \text{if $p = (1- \gamma_{i}) \pi_{j_{i}} + \gamma_{i} \pi_{j_{i} + 1}$}\\
   & \text{$0 < \gamma_{i} < 1$, $j_{i} = 1,\ldots, k - 1$, $i = 1,\ldots,n$,}\\
y_{j_{k}}, & \text{if $p > \pi_{j_{k}}$.}\\
\end{cases}
\end{equation*}
The dependence of $\pi_{j_{i}}$, $y_{j_{i}}$ and $\gamma_{i}$ on $i$ is the consequence of \emph{conditioning} on $x_{i}$. Clearly, one would need to calculate all the mid-probabilities $\pi_{j}$, $j = 1, \ldots,k$, because the value of $H_{Y|X = x_{i}}(p)$, $i = 1, \ldots,n$, depends on where $p$ lies with respect to the $\pi_{j}$'s. However, if one starts from the implicit equation $p = \hat{G}_{Y|X}^{c}\left(\eta(p)|x\right)$ as we did, it becomes easier to work out the asymptotic properties of the estimator using the objective function (2.7).

\subsection{Prediction of conditional mid-quantiles and recovery of ordinary quantiles}
\label{sec:2.4}
So far, we have focused our attention on the $\beta(p)$, which is and remains the primary goal of inference in our modeling approach. However, it is immediate to obtain a prediction of the conditional mid-quantiles of $Y$ conditional on $X=x$ with $\hat{H}_{Y|X}(p) = h^{-1}\left(x\tp\hat{\beta}(p)\right)$. The approximate sampling distribution of $\hat{H}_{Y|X}(p)$ is given in Corollary~\ref{theo:3}.

It might also be of some interest to know that we are able to recover ordinary conditional quantiles from the conditional mid-quantiles. As this result represents a by-product of our models, we do not place it in direct competition with alternative approaches \citep{Chernozhukov2019}, which may well be preferable to ours since ours is not specifically designed for such a purpose. The goal is to relabel $\hat{H}_{Y|X}(p)$ to obtain an estimate of the (ordinary) conditional quantiles $Q_{Y|X}(p) \equiv \inf\{y \in \mathbb{R}: F_{Y|X}(y)\geq p\}$. Let us denote such an estimate by $\tilde{Q}_{Y|X}(p)$. This can be achieved in few simple steps. Since the distribution function $F_{Y|X}$ has been already estimated in order to fit the mid-quantile regression model, it is sufficient to plug $x$ into $\hat{F}_{Y|X}(y|x)$. The latter is used to obtain $\hat{G}_{Y|X}(y|x)$ as described in the previous section. Successively, we identify the index $j$ such that $\hat{G}_{Y|X}(z_{j}|x) \leq p < \hat{G}_{Y|X}(z_{j+1}|x)$. If $Y$ is a variable whose values are irregularly spaced, then one would naturally take $\tilde{Q}_{Y|X}(p) = z_{j + 1}$ if $p > \hat{F}_{Y|X}(z_{j}|x)$ or $\tilde{Q}_{Y|X}(p) = z_{j}$ otherwise, with the understanding that $\tilde{Q}_{Y|X}(p) = z_{1}$ if $p < \hat{F}_{Y|X}(z_{1}|x)$. If, on the other hand, $Y$ is a variable with equally-spaced values (e.g., a count), then we can take $\tilde{Q}_{Y|X}(p) = \lceil \hat{H}_{Y|X}(p) \rceil$ if $p > \hat{F}_{Y|X}(z_{j}|x)$ or $\tilde{Q}_{Y|X}(p) = \lfloor \hat{H}_{Y|X}(p) \rfloor$ otherwise, with the understanding that $\tilde{Q}_{Y|X}(p) = \lceil \hat{H}_{Y|X}(p) \rceil$ if $p < \hat{F}_{Y|X}(z_{1}|x)$. A demonstration is given in Figure~\ref{fig:7}. We do not pursue the calculation of standard errors for $\tilde{Q}_{Y|X}(p)$, which can be derived from the distribution of $\hat{H}_{Y|X}(p)$ (Corollary~\ref{theo:3}), as it is not within our main interests.

\subsection{Theoretical results}
\label{sec:2.5}

We first show consistency of $\hat{\beta}(p)$ under general assumptions. We then provide its asymptotic distribution. Here, we assume that the conditional CDF estimator of \cite{Li2008} is used to obtain $\hat F_{Y|X}(y|x)$. The identity matrix of order $m$ will be denoted by $I_{m}$. The proofs of the theorems in this section are given in Appendix~\ref{sec:A}.

\begin{thm}
\label{theo:1}
Generate $n$ independent conditional responses from a discrete distribution with parameters satisfying \eqref{eqn:4}. Assume that the marginal density of the continuous covariates in $X$ is strictly positive and that $0<F_{Y|X}(y|x)<1$. Assume also that the kernel $K_{\lambda}(X,x)$ in \eqref{eqn:6} is symmetric, bounded, and compactly supported, and that $n\prod_j \lambda_j \to \infty$, while $\lambda_j\to 0$ for all $j=1,\ldots,q$.
For a fixed $p\in (0,1)$, let $\hat\beta(p)$ denote the solution in \eqref{eqn:8} and let $\beta^{*}(p)$ be its population counterpart.

Then, as $n \to \infty$, $\sup_z \left|\hat G^{c}_{Y|X}(z|x) - G^{c}_{Y|X}(z|x)\right| \to 0$. Additionally, $\|\hat\beta(p)-\beta^{*}(p)\| \to 0$ and $\left|\hat{H}_{h(Y)|X}(p)-H_{h(Y)|X}(p)\right| \to 0$, for all $p \in (0,1)$.
\end{thm}

\begin{thm}
\label{theo:2}
In addition to the assumptions in Theorem \ref{theo:1}, assume that\\ $\hat{G}_{Y|X}^{c}\left(h^{-1}(x\tp\beta)|x\right)$ is differentiable with respect to $\beta$. Assume also that the design matrix is full rank, that $\lim_{n} n^{-1}XX\tp$ exists and is a positive-definite matrix. Finally, assume that $\sqrt{n\prod_j \lambda_j}(\sum_j\lambda_j)^2 = O(1)$. Then,
\begin{equation*}
V(\beta^{*}(p))^{-1/2}\sqrt{n\prod_j \lambda_j}(\hat\beta(p)-\beta^{*}(p)) \to N(0,I_q) \qquad \mbox{in distribution},
\end{equation*}
where $V(\beta^{*}(p)) = J(\beta^{*}(p))^{-1}D(\beta^{*}(p))J(\beta^{*}(p))^{-1}$,\\ $J(\beta^{*})=E\left\{\nabla^2_{\beta} \psi_n(\beta;p)\Big\rvert_{\beta = \beta^{*}}\right\}$, and $D(\beta^{*}) = {\rm Var}\left\{\sqrt{n\prod_j\lambda_j} \nabla_{\beta} \psi_n(\beta^{*};p) \Big\rvert_{\beta = \beta^{*}}\right\}$.
\end{thm}

\begin{corollary}
\label{theo:3}
The approximate sampling distribution of $\hat{H}_{Y|X}(p) = h^{-1}\left(x\tp\hat{\beta}(p)\right)$ follows from the results in Theorems \ref{theo:1} and \ref{theo:2}. Let $\Sigma_{p} = V(\beta^{*}(p))(n\prod_j \lambda_j)^{-1}$ and let $\delta(b) = \nabla_{\beta}h^{-1}\left(x\tp\beta\right)\Big\rvert_{\beta = b}$. Then for large $n$, the distribution of $\hat{H}_{Y|X}(p)$ is approximately normal with mean $h^{-1}\left(x\tp\beta^{*}(p)\right)$ and variance $\delta(\beta^{*}(p))\tp \Sigma_{p} \delta(\beta^{*}(p))$.
\end{corollary}

\section{Results}
\label{sec:3}
In this section, we illustrate the performance of mid-quantile regression on both simulated and real data. The application concerns prescription drugs use in the United States (US).

\subsection{Simulation study}
\label{sec:3.1}

Data were generated according to six distinct models. The first four had homoscedastic discrete uniform, heteroscedastic discrete uniform, Poisson, and Bernoulli errors, respectively. Each of these four models was considered with either one discrete covariate (1a, 2a, 3a, and 4a), or with two continuous covariates (1b, 2b, 3b, and 4b). The fifth model was defined as the ratio of two Poisson random variables. Lastly, the response variable for the sixth model was randomly sampled (with replacement) from the empirical (marginal) distribution of prescription drugs analyzed in Section~\ref{sec:3.2} and depicted in Figure~\ref{fig:5}. While the responses generated with the first four models are equally-spaced integers from standard, \emph{regular} distributions, the last two provide instances of non-standard features like irregularly-spaced values, fractional values, zero-excess, and outliers. In symbols, data were generated as follows:
\begin{itemize}
\item[(1a)] $Y = \lfloor 1 + 2w \rfloor + \epsilon$, where $w \sim \mathrm{DU}(0, 5)$ and $\epsilon \sim \mathrm{DU}(1, 10)$;
\item[(1b)] $Y = \lfloor 1 + 2w_{1} + w_{2} \rfloor + \epsilon$, where $w_{1} \sim \mathrm{U}(0, 5)$,  $w_{2} \sim 1/3\chi^{2}_{3}$, and $\epsilon \sim \mathrm{DU}(1, 10)$;
\item[(2a)] $Y = \lfloor 1 + 2w \rfloor + \lfloor w + 1\rfloor\epsilon$, where $w \sim \mathrm{DU}(0, 5)$ and $\epsilon \sim \mathrm{DU}(1, 10)$;
\item[(2b)] $Y = \lfloor 1 + 2w_{1} + w_{2}\rfloor + \lfloor w_{1} + 1\rfloor\epsilon$, where $w_{1} \sim \mathrm{U}(0, 5)$,  $w_{2} \sim 1/3\chi^{2}_{3}$, and $\epsilon \sim \mathrm{DU}(1, 10)$;
\item[(3a)] $Y = \epsilon$, where $\epsilon \sim \mathrm{Poisson}(\mu)$, $\mu = \exp(0.5 + 2w)$, and $w \sim \mathrm{DU}(1, 3)$;
\item[(3b)] as in scenario (3a) with $\mu = \exp(0.5 + 2w_{1} + 0.3w_{2})$, $w_{1} \sim \mathrm{U}(1, 3)$, and $w_{2} \sim 1/3\chi^{2}_{3}$;
\item[(4a)] $Y = \epsilon$, where $\epsilon \sim \mathrm{Bernoulli}(\mu)$, $\mu = 1/\left[1+\exp\{-(3 + w)\}\right]$, and $w \sim \mathrm{DU}(0, 5)$;
\item[(4b)] as in scenario (4a) with $\mu = 1/\left[1+\exp\{-(3 + w_{1} + w_{2})\}\right]$, $w_{1} \sim \mathrm{U}(0, 5)$, and $w_{2} \sim 1/3\chi^{2}_{3}$;
\item[(5)] $Y = \epsilon_{1}/(\epsilon_{2} + 1)$, where $\epsilon_{h} \sim \mathrm{Poisson}(\mu)$, $h = 1, 2$, $\mu = \exp(0.5 + 1w)$, and $w \sim \mathrm{DU}(1, 3)$;
\item[(6)] $Y$ is a discrete variable with values in $S_{Y} = \{0, 1, 4, 9, 16, 25, 36, 49, 64,\\ 81, 100, 121, 144, 169, 196, 225, 256, 289, 361, 400\}$ and corresponding probabilities $\{5.0\times 10^{-1}, 1.8\times 10^{-1}, 9.6\times 10^{-2}, 5.4\times 10^{-2}, 4.9\times 10^{-2}, 3.4\times 10^{-2}, 2.4\times 10^{-2}, 2.0\times 10^{-2}, 1.1\times 10^{-2}, 8.9\times 10^{-3}, 5.1\times 10^{-3}, 2.8\times 10^{-3}, 3.3\times 10^{-3}, 2.3\times 10^{-3}, 2.3\times 10^{-3}, 9.3\times 10^{-4}, 9.3\times 10^{-4}, 1.9\times 10^{-3}, 4.7\times 10^{-4}, 4.7\times 10^{-4}\}$, and $w \sim \mathrm{DU}(1, 3)$;
\end{itemize}
where $\mathrm{DU}(a,b)$ and $\mathrm{U}(a,b)$ denote random variables with, respectively, discrete and continuous uniform distribution on $(a,b)$. Samples $(y_{i}, w_{i})$ of size $n \in \{100,500,1000\}$ were independently drawn from each model for $R = 1000$ replications. We then fitted the linear mid-quantile model $H_{Y|X}(p) = x\tp\beta(p)$ with data generated under models 1, 2, 5, and 6; the log-linear mid-quantile model $H_{Y|X}(p) = \exp\{x\tp\beta(p)\}$ with data generated under model 3; and the logistic mid-quantile model $H_{Y|X}(p) = [1 + \exp\{-x\tp\beta(p)\}]^{-1}$ with data generated under model 4. (Note, however, that the sampling probabilities under model 6 do not depend on $w$.) All models were estimated for 7 deciles, $p \in \{0.2, 0.3, \ldots, 0.8\}$, except the logistic model, which was estimated for the median only.

Let $H_{i}(p_{k}) \equiv H_{Y_{i}|X_{i}}(p_{k})$ denote the true mid-quantile at level $p_{k}$ for a given $x_{i} = (1, w_{i})\tp$ under any of the data-generating models defined above and $\hat{H}^{(r)}_{i}(p_{k})$ be the corresponding estimate for replication $r$. We assessed the performance of the proposed methods in terms of average bias and root mean squared error (RMSE) of the mid-QF, i.e.
\begin{equation*}
\frac{1}{R} \sum_{r = 1}^{R} \left\{n^{-1}\sum_{i = 1}^{n} \hat{H}^{(r)}_{i}(p_{k}) - H_{i}(p_{k})\right\}
\end{equation*}
and
\begin{equation*}
\left[\frac{1}{R} \sum_{r = 1}^{R} \left\{n^{-1}\sum_{i = 1}^{n} \left(\hat{H}^{(r)}_{i}(p_{k}) - H_{i}(p_{k})\right)^{2}\right\}\right]^{\frac{1}{2}},
\end{equation*}
respectively. We also report the average true mid-quantiles at $n = 1000$
\[
\bar{H}(p_{k}) = n^{-1}\sum_{i = 1}^{n} H_{i}(p_{k})
\]
as a term of comparison for assessing the relative magnitude of the bias. Finally, we calculated $95\%$ confidence intervals to assess coverage of the slope parameter in mid-quantile models for $p \in \{0.3, 0.5, 0.7\}$ when data were generated under scenarios 1a, 2a, and 3a. The corresponding standard errors were computed based on expression \eqref{eqn:10}.

Estimated bias and RMSE of the proposed estimator are shown in Tables~\ref{tab:1}-\ref{tab:6} for scenarios 1a, 2a, 3a, 4a, 5, and 6, and in Tables \ref{tab:B1}-\ref{tab:B4} (Appendix~\ref{sec:B}) for scenarios 1b, 2b, 3b, and 4b. The bias was, in general, small, never exceeding $2.3\%$ of the average mid-quantile for the homoscedastic discrete uniform model (1a and 1b), $3.3\%$ for the heteroscedastic discrete uniform model (2a and 2b), $0.6\%$ for the Poisson model with a discrete covariate (3a), and $2.5\%$ for the Poisson ratio model (5). The estimated bias and RMSE of the proposed estimator for the Bernoulli model (4a and 4b) were extremely small at all sample sizes. In contrast the bias was relatively higher (up to $11\%$ of the average mid-quantile) in the Poisson scenario with continuous covariates (3b), although this issue was limited to the tail quantiles at smaller sample sizes. The bias and RMSE for model (6) were notable at smaller sample sizes. In particular, the bias was larger at $p = 0.6$. This can be explained by examining the mid-CDF in Figure~\ref{fig:5}. The function goes from $0.25$ at $y = 0$ to just shy of $0.6$ at $y = 1$. Conditioning on $w$ skews the distribution of $\hat{\beta}(0.6)$ since $w$ is unrelated to the response (further investigation revealed that, at $p = 0.6$, the estimator is median-unbiased). In general, both bias and RMSE decreased with $n$ at approximately the expected rate for all six models.

\begin{figure}[ht]
\centering
\includegraphics[scale = 0.30]{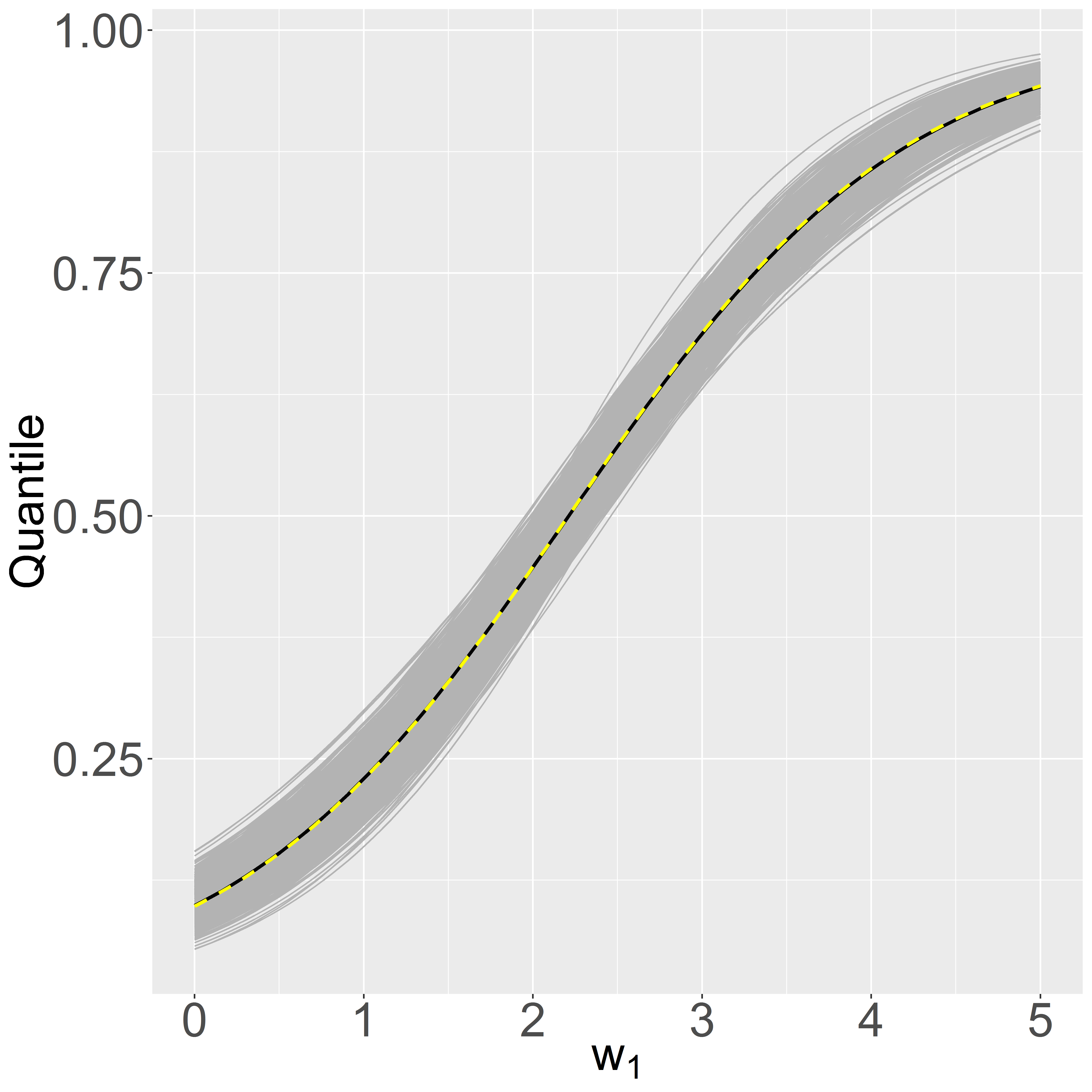}
\caption{True quantile function (black line), estimated conditional mid-quantile functions for all replications (grey lines), and average estimated conditional mid-quantile function (dashed yellow) for the Bernoulli simulated scenario with continuous covariates and $n = 1000$.}
\label{fig:4}
\end{figure}

To appreciate how these results translate into model fitting, the estimated conditional mid-quantiles from all replications and the average estimated conditional mid-quantiles ($n = 1000$) are shown in Figure~\ref{fig:3} for scenarios 1b, 2b, and 3b, and in Figure~\ref{fig:4} for scenario 4b. All mid-quantiles are plotted as functions of $w_{1}$, with $w_{2}$ set equal to the median of $1/3\chi^{2}_{3}$.

The observed coverage at the nominal $95\%$ confidence level for the slope in selected scenarios is given in Table~\ref{tab:7}. The results are in general accurate, although frequencies are occasionally slightly away from the nominal level. This is not surprising since the sample estimator of \eqref{eqn:10} relies on the Huber-White estimator and on several approximations.

It would be remiss of us not to make a contrast between our proposed estimator and existing alternatives. The estimator developed by \cite{Machado2005} (hereinafter referred to as MSS) is a natural candidate. However, comparison is inevitably restricted: first of all, neither the `true' coefficients nor the population quantiles underlying mid-quantile and jittering-based estimation are necessarily the same quantities. Indeed, the quantiles modeled by MSS are defined as $Q_{Z|X}(p) = Q_{Y|X}(p) + \frac{p - F_{Y|X}\{Q_{Y|X}(p) - 1\}}{m_{Y|X}(Q_{Y|X}(p))}$, where $Z = Y + U$ and $U$ is uniformly distributed on $[0,1)$. The jittered quantiles $Q_{Z|X}(p)$ dominate the true quantiles since $F_{Y|X}\{Q_{Y|X}(p) - 1\} \leq p$, uniformly over $p$. In contrast, mid-quantiles interpolate the true quantiles. In addition, the comparison must be restricted to when the response is a count or ordinal variable, as these are required by jittering-based estimation. For these reasons, we considered only Poisson data as in scenario 3a, in which case we expected the two estimators to target the same functional relationship (i.e., slope). The mean and variance of the estimates using our estimator (MIDQR), as well as the ratio of means and variances comparing the MSS estimator relative to MIDQR are given in Table~\ref{tab:8}. The two estimators gave similar estimates of the slope with MSS:MIDQR ratios of the means close to 1 across quantiles and sample sizes. Our estimator was generally more efficient (that is, with MSS:MIDQR ratios of the variances greater than 1) consistently for $0.2 < p < 0.8$. The MSS estimator showed a faster convergence rate at $p = 0.2$ and $p = 0.8$. However, this advantage withered away as the sample size increased. At $p=0.2$, the ratio of the variances went from 0.878 ($n = 100$) to 0.978 ($n = 1000$), while at $p = 0.8$, it went from 0.939 ($n = 100$) to 1.072 ($n = 1000$). That is, for $n = 1000$, the two estimators performed similarly on the tails. We conclude this section by remarking that these results are clearly not generalizable.

\clearpage


\begin{table}
\caption{Bias and root mean squared error (RMSE) of predicted quantiles for data generated using the homoscedastic discrete uniform model (1a). \label{tab:1}}
\begin{tabular}{lrrrrrrr}
\hline
 & \multicolumn{2}{c}{$n = 100$} & \multicolumn{2}{c}{$n = 500$} & \multicolumn{2}{c}{$n = 1000$} & \\
$p$ & \multicolumn{1}{c}{Bias} & \multicolumn{1}{c}{RMSE} & \multicolumn{1}{c}{Bias} & \multicolumn{1}{c}{RMSE} & \multicolumn{1}{c}{Bias} & \multicolumn{1}{c}{RMSE} & \multicolumn{1}{c}{$\bar{H}$}\\
\hline
0.2 & 0.015 & 0.508 & $-$0.023 & 0.221 & $-$0.006 & 0.153 & 8.494 \\
  0.3 & 0.097 & 0.548 & 0.004 & 0.246 & 0.011 & 0.172 & 9.494 \\
  0.4 & 0.113 & 0.574 & 0.008 & 0.263 & 0.016 & 0.182 & 10.494 \\
  0.5 & 0.121 & 0.587 & 0.010 & 0.271 & 0.016 & 0.186 & 11.494 \\
  0.6 & 0.121 & 0.579 & 0.012 & 0.269 & 0.013 & 0.183 & 12.494 \\
  0.7 & 0.135 & 0.550 & 0.014 & 0.251 & 0.012 & 0.167 & 13.494 \\
  0.8 & 0.211 & 0.532 & 0.045 & 0.223 & 0.028 & 0.146 & 14.494\\
\hline
\end{tabular}
\end{table}

\begin{table}
\caption{Bias and root mean squared error (RMSE) of predicted quantiles for data generated using the heteroscedastic discrete uniform model (2a). \label{tab:2}}
\begin{tabular}{lrrrrrrr}
\hline
 & \multicolumn{2}{c}{$n = 100$} & \multicolumn{2}{c}{$n = 500$} & \multicolumn{2}{c}{$n = 1000$} & \\
$p$ & \multicolumn{1}{c}{Bias} & \multicolumn{1}{c}{RMSE} & \multicolumn{1}{c}{Bias} & \multicolumn{1}{c}{RMSE} & \multicolumn{1}{c}{Bias} & \multicolumn{1}{c}{RMSE} & \multicolumn{1}{c}{$\bar{H}$}\\
\hline
0.2 & $-$0.417 & 1.689 & $-$0.281 & 1.056 & $-$0.071 & 0.791 & 14.737 \\
  0.3 & $-$0.482 & 1.940 & $-$0.389 & 1.108 & $-$0.229 & 0.856 & 18.234 \\
  0.4 & $-$0.397 & 2.038 & $-$0.181 & 1.071 & 0.048 & 0.817 & 21.731 \\
  0.5 & $-$0.444 & 2.100 & $-$0.418 & 1.068 & $-$0.297 & 0.810 & 25.228 \\
  0.6 & $-$0.308 & 2.134 & $-$0.322 & 1.138 & $-$0.290 & 0.887 & 28.725 \\
  0.7 & $-$0.073 & 1.933 & $-$0.252 & 0.967 & $-$0.168 & 0.724 & 32.222 \\
  0.8 & 0.484 & 1.864 & 0.273 & 0.817 & 0.259 & 0.594 & 35.719\\
\hline
\end{tabular}
\end{table}

\begin{table}
\caption{Bias and root mean squared error (RMSE) of predicted quantiles for data generated using the Poisson model (3a). \label{tab:3}}
\begin{tabular}{lrrrrrrr}
\hline
 & \multicolumn{2}{c}{$n = 100$} & \multicolumn{2}{c}{$n = 500$} & \multicolumn{2}{c}{$n = 1000$} & \\
$p$ & \multicolumn{1}{c}{Bias} & \multicolumn{1}{c}{RMSE} & \multicolumn{1}{c}{Bias} & \multicolumn{1}{c}{RMSE} & \multicolumn{1}{c}{Bias} & \multicolumn{1}{c}{RMSE} & \multicolumn{1}{c}{$\bar{H}$}\\
\hline
0.2 & 0.232 & 3.369 & 0.190 & 1.848 & 0.226 & 1.552 & 243.938 \\
  0.3 & 0.711 & 2.921 & 0.482 & 1.476 & 0.492 & 1.199 & 247.933 \\
  0.4 & 0.554 & 2.794 & 0.340 & 1.358 & 0.377 & 1.037 & 251.596 \\
  0.5 & 0.862 & 2.892 & 0.635 & 1.434 & 0.642 & 1.144 & 254.593 \\
  0.6 & 0.891 & 2.983 & 0.605 & 1.454 & 0.610 & 1.122 & 257.921 \\
  0.7 & 1.265 & 3.462 & 0.856 & 1.842 & 0.844 & 1.504 & 261.251 \\
  0.8 & 1.580 & 4.288 & 0.888 & 2.388 & 0.827 & 2.050 & 265.580\\
\hline
\end{tabular}
\end{table}

\begin{table}[ht]
\caption{Bias and root mean squared error (RMSE) of predicted quantiles for data generated using the Bernoulli model (4a). \label{tab:4}}
\begin{tabular}{lrrrrrrr}
\hline
 & \multicolumn{2}{c}{$n = 100$} & \multicolumn{2}{c}{$n = 500$} & \multicolumn{2}{c}{$n = 1000$} & \\
$p$ & \multicolumn{1}{c}{Bias} & \multicolumn{1}{c}{RMSE} & \multicolumn{1}{c}{Bias} & \multicolumn{1}{c}{RMSE} & \multicolumn{1}{c}{Bias} & \multicolumn{1}{c}{RMSE} & \multicolumn{1}{c}{$\bar{H}$}\\
\hline
0.5 & 0.001 & 0.047 & 0.001 & 0.022 & $-$0.000 & 0.015 & 0.424 \\
   \hline
\end{tabular}
\end{table}

\begin{table}[ht]
\caption{Bias and root mean squared error (RMSE) of predicted quantiles for data generated using the Poisson ratio model (5). \label{tab:5}}
\begin{tabular}{lrrrrrrr}
\hline
 & \multicolumn{2}{c}{$n = 100$} & \multicolumn{2}{c}{$n = 500$} & \multicolumn{2}{c}{$n = 1000$} & \\
$p$ & \multicolumn{1}{c}{Bias} & \multicolumn{1}{c}{RMSE} & \multicolumn{1}{c}{Bias} & \multicolumn{1}{c}{RMSE} & \multicolumn{1}{c}{Bias} & \multicolumn{1}{c}{RMSE} & \multicolumn{1}{c}{$\bar{H}$}\\
\hline
0.2 & $-$0.005 & 0.063 & $-$0.010 & 0.031 & $-$0.008 & 0.024 & 0.643 \\
  0.3 & $-$0.008 & 0.065 & $-$0.012 & 0.032 & $-$0.011 & 0.024 & 0.736 \\
  0.4 & $-$0.007 & 0.069 & $-$0.011 & 0.033 & $-$0.010 & 0.025 & 0.821 \\
  0.5 & $-$0.000 & 0.076 & $-$0.006 & 0.033 & $-$0.006 & 0.023 & 0.905 \\
  0.6 & $-$0.009 & 0.087 & $-$0.008 & 0.041 & $-$0.005 & 0.031 & 1.015 \\
  0.7 & $-$0.018 & 0.107 & $-$0.028 & 0.063 & $-$0.029 & 0.055 & 1.144 \\
  0.8 & $-$0.017 & 0.134 & $-$0.022 & 0.064 & $-$0.021 & 0.050 & 1.304 \\
   \hline
\end{tabular}
\end{table}

\begin{table}[ht]
\caption{Bias and root mean squared error (RMSE) of predicted quantiles for data generated using the NHANES empirical distribution model (6). \label{tab:6}}
\begin{tabular}{lrrrrrrr}
\hline
 & \multicolumn{2}{c}{$n = 100$} & \multicolumn{2}{c}{$n = 500$} & \multicolumn{2}{c}{$n = 1000$} & \\
$p$ & \multicolumn{1}{c}{Bias} & \multicolumn{1}{c}{RMSE} & \multicolumn{1}{c}{Bias} & \multicolumn{1}{c}{RMSE} & \multicolumn{1}{c}{Bias} & \multicolumn{1}{c}{RMSE} & \multicolumn{1}{c}{$\bar{H}$}\\
\hline
0.2 & 0.003 & 0.004 & 0.000 & 0.000 & 0.000 & 0.000 & 0.000 \\
  0.3 & 0.018 & 0.078 & 0.002 & 0.033 & 0.001 & 0.024 & 0.143 \\
  0.4 & 0.023 & 0.098 & 0.003 & 0.039 & 0.002 & 0.029 & 0.436 \\
  0.5 & 0.060 & 0.152 & 0.004 & 0.047 & 0.002 & 0.034 & 0.729 \\
  0.6 & 0.417 & 0.654 & 0.111 & 0.287 & 0.070 & 0.217 & 1.170 \\
  0.7 & 0.492 & 1.242 & 0.054 & 0.400 & 0.008 & 0.273 & 3.348 \\
  0.8 & 0.979 & 3.151 & 0.286 & 1.292 & 0.129 & 0.908 & 8.651 \\
   \hline
\end{tabular}
\end{table}

\begin{table}
\caption{Observed coverage at the nominal $95\%$ confidence level for the slope in mid-quantiles models for data generated using the homoscedastic discrete uniform (1a), heteroscedastic discrete uniform (2a), and Poisson (3a) models. \label{tab:7}}
\begin{tabular}{lrrrrrrrrr}
\hline
 &  \multicolumn{3}{c}{1a} &  \multicolumn{3}{c}{2a} &  \multicolumn{3}{c}{3a}\\
$p$ & \multicolumn{1}{c}{100} & \multicolumn{1}{c}{500} & \multicolumn{1}{c}{1000} & \multicolumn{1}{c}{100} & \multicolumn{1}{c}{500} & \multicolumn{1}{c}{1000} & \multicolumn{1}{c}{100} & \multicolumn{1}{c}{500} & \multicolumn{1}{c}{1000}\\
\hline
0.3 & 97.70 & 97.70 & 97.90 & 96.60 & 95.90 & 93.30 & 93.70 & 94.59 & 95.09 \\
  0.5 & 95.90 & 94.90 & 96.10 & 94.60 & 94.60 & 92.50 & 93.90 & 95.19 & 95.09 \\
  0.7 & 98.30 & 96.70 & 98.50 & 96.00 & 96.00 & 96.80 & 97.60 & 97.49 & 96.99\\
\hline
\end{tabular}
\end{table}

\begin{table}
\caption{Mean and variance ($\times 1000$) of the slope's estimates using the proposed approach (MIDQR), and ratio of mean and variance of the estimates using \citeauthor{Machado2005}'s (\citeyear{Machado2005}) estimator (MSS) compared to MIDQR for data generated using the Poisson model (3a). A MSS:MIDQR ratio greater than 1 indicates a larger MSS value. \label{tab:8}}
\begin{tabular}{lrrrrrrrrrrrr}
\hline
  &  \multicolumn{4}{c}{$n=100$} &  \multicolumn{4}{c}{$n=500$} &  \multicolumn{4}{c}{$n=1000$}\\
  & \multicolumn{2}{c}{MIDQR} & \multicolumn{2}{c}{MSS:MIDQR} & \multicolumn{2}{c}{MIDQR} & \multicolumn{2}{c}{MSS:MIDQR}& \multicolumn{2}{c}{MIDQR} & \multicolumn{2}{c}{MSS:MIDQR}\\
$p$ & \multicolumn{1}{c}{Mean} & \multicolumn{1}{c}{Var.} & \multicolumn{1}{c}{Mean} &  \multicolumn{1}{c}{Var.}& \multicolumn{1}{c}{Mean} & \multicolumn{1}{c}{Var.} & \multicolumn{1}{c}{Mean} & \multicolumn{1}{c}{Var.} & \multicolumn{1}{c}{Mean} & \multicolumn{1}{c}{Var.} & \multicolumn{1}{c}{Mean} & \multicolumn{1}{c}{Var.}\\
\hline
0.2 & 2.094 & 0.997 & 1.000 & 0.878 & 2.091 & 0.189 & 1.000 & 0.895 & 2.091 & 0.089 & 1.001 & 0.978 \\
  0.3 & 2.051 & 0.489 & 1.003 & 1.316 & 2.053 & 0.097 & 1.002 & 1.312 & 2.053 & 0.055 & 1.002 & 1.214 \\
  0.4 & 2.022 & 0.409 & 1.004 & 1.384 & 2.025 & 0.081 & 1.002 & 1.307 & 2.026 & 0.043 & 1.002 & 1.364 \\
  0.5 & 1.997 & 0.386 & 1.004 & 1.442 & 2.002 & 0.073 & 1.002 & 1.257 & 2.003 & 0.039 & 1.001 & 1.308 \\
  0.6 & 1.974 & 0.372 & 1.004 & 1.357 & 1.979 & 0.072 & 1.001 & 1.332 & 1.980 & 0.037 & 1.001 & 1.399 \\
  0.7 & 1.947 & 0.423 & 1.005 & 1.250 & 1.955 & 0.081 & 1.001 & 1.220 & 1.956 & 0.040 & 1.000 & 1.387 \\
  0.8 & 1.913 & 0.578 & 1.008 & 0.939 & 1.924 & 0.114 & 1.002 & 1.025 & 1.926 & 0.058 & 1.001 & 1.072\\
\hline
\end{tabular}
\end{table}

\begin{landscape}
\centering
\begin{figure}
\includegraphics[scale = 0.4]{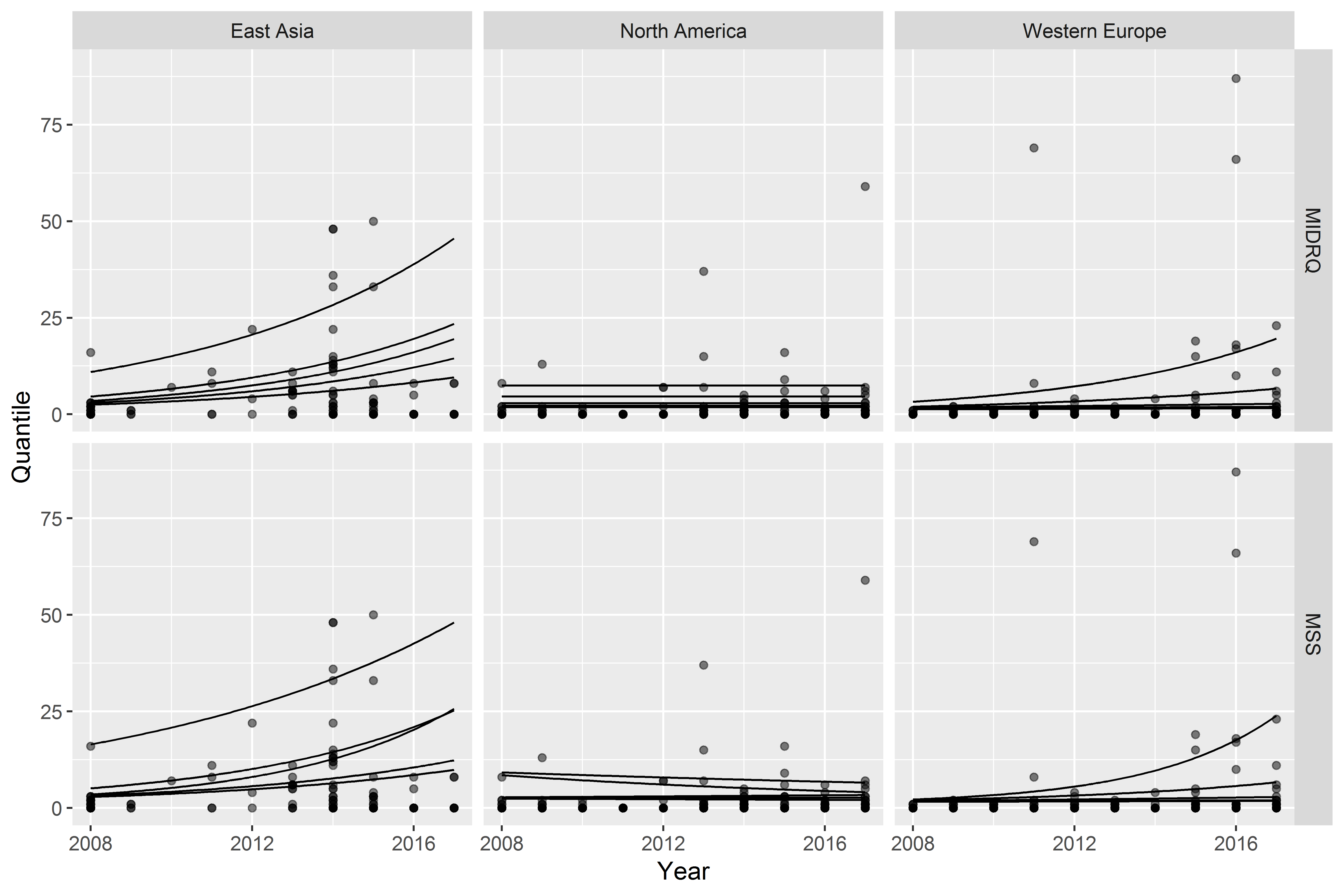}
\caption{True quantile function (black line), estimated conditional mid-quantile functions for all replications (grey lines), and average estimated conditional mid-quantile function (dashed yellow) for three simulated scenarios with continuous covariates and $n = 1000$.}
\label{fig:3}
\end{figure}
\end{landscape}

\subsection{Prescription drugs}
\label{sec:3.2}

In this section we illustrate an application of mid-quantile regression using data on prescription medications from the National Health and Nutrition Examination Survey (NHANES) \citep{NHANES}. The US is the worldwide leader in per capita prescription drug spending \citep{KFF} and its pharmaceutical market represents a major economic sector worth hundreds of billions of dollars. In a recent quantile regression analysis of NHANES data, \cite{Hong2019} found a higher opioid use (morphine milligram equivalent) in adults with longstanding physical disability and those with inflammatory conditions as compared to individuals with other conditions. Differences were markedly larger at the 75th and 95th percentiles than those at lower percentiles. In the context of medications use, a higher percentile can be interpreted as an index of diminished health, lower quality of life, and higher financial burden. A quantile regression analysis of prescription medications use is therefore of both public health and health economics interest.

\begin{figure}[t!]
\centering
\includegraphics[scale = 0.55]{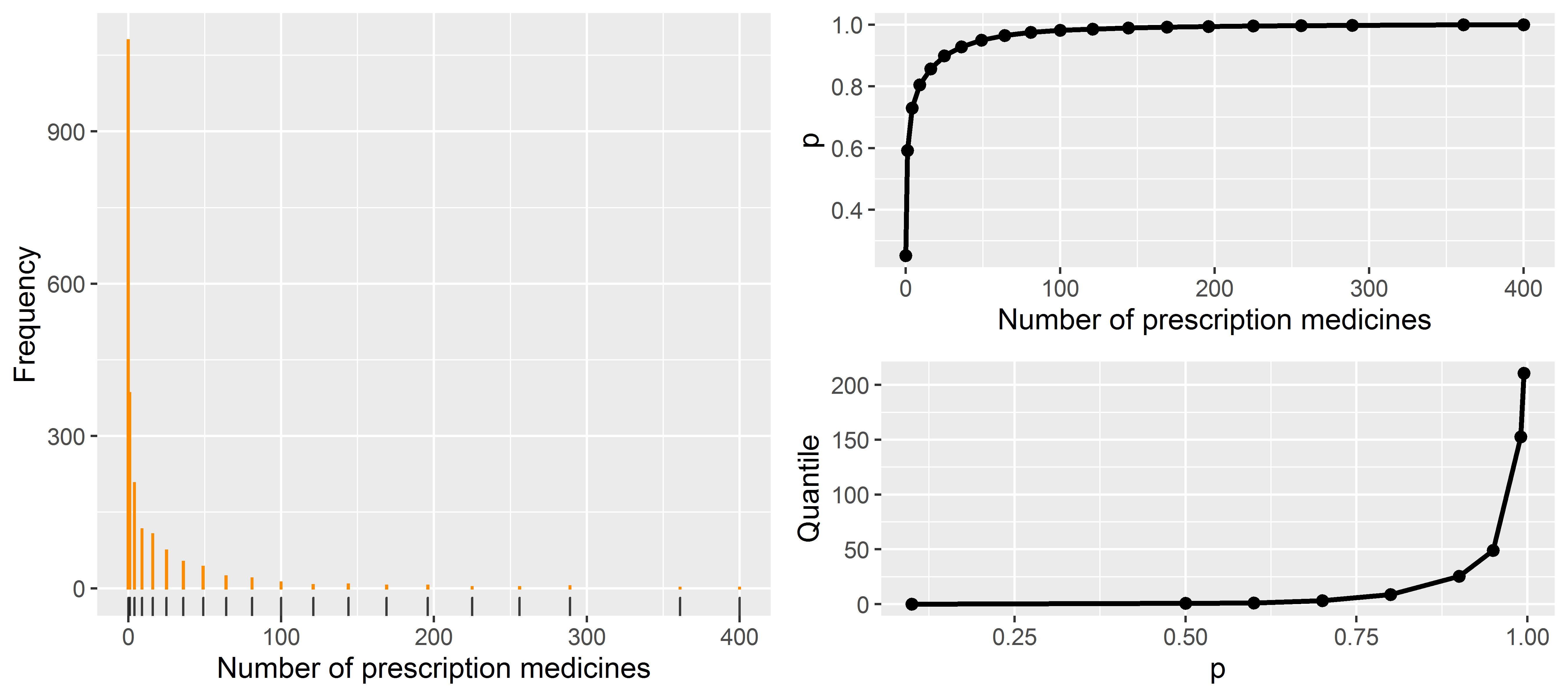}
\caption{Number of prescription medicines using the National Health and Nutrition Examination Survey data, 2015-2016. Left: frequency bar plot with rug plot. Right: estimated mid-cumulative distribution function $\hat{G}^{c}(y)$ with filled circles marking points $y \in \{0, 1, 4, 9, 16, 25, 36, 49, 64, 81, 100, 121, 144, 169, 196, 225, 256, 289, 361, 400\}$ (top) and estimated mid-quantile function $\hat{H}(p)$ with filled circles marking points $p \in \{0.1, 0.5, 0.6, 0.7, 0.8, 0.9, 0.95, 0.99, 0.995\}$ (bottom). \label{fig:5}}
\end{figure}

\begin{table}[!h]
\caption{Descriptive summary for the National Health and Nutrition Examination Survey data, 2015-2016. \label{tab:9}}
\begin{tabular}{lrrrrr}
\hline
\emph{Variable} & Minimum & \multicolumn{1}{c}{First quartile} & \multicolumn{1}{c}{Median} & \multicolumn{1}{c}{Third quartile} & Maximum\\
\hline
Prescription medicines & & & & & \\
\multicolumn{1}{r}{Overall} & 0.0 & 0.0 & 2.0 & 8.1 & 400.0 \\
\multicolumn{1}{r}{Female} & 0.0 & 0.2 & 2.5 & 9.5 & 400.0\\
\multicolumn{1}{r}{Male} & 0.0 & 0.0 & 2.0 & 6.7 & 289.0\\
Age (years) & 18.0 & 31.0 & 40.3 & 49.1 & 65.0 \\
BMI (kg/m$^2$) & 14.3 & 24.6 & 28.0 & 31.5 & 57.2\\
\hline
\emph{Variable} & \multicolumn{1}{c}{Frequency} & \multicolumn{1}{c}{Proportion (\%)} &  & &\\
\hline
Sex (female) & 1061 & 49.4 &  &  &  \\
Health status  & 1723 & 80.3 &  &  &  \\
(good or excellent) & &  &  &  &  \\
Smoking status (no) & 1219 & 56.8 &  &  &  \\
Alcohol use (no) & 547 & 25.5 &  &  & \\
\hline
\end{tabular}
\end{table}

We abstracted data ($n=9,\!971$) on number of prescription medicines taken from the 2015-2016 Dietary Supplement and Prescription Medication section of the Sample Person Questionnaire. We also obtained information on sex, age (years), perceived health status, smoking status (`smoked at least 100 cigarettes in life'), alcohol use (`had at least 12 alcohol drinks in 1 year'), weight (kg), height (m), and race. Before carrying out the analysis, we removed the effect of NHANES oversampling by first restricting the dataset to all observations for White persons (about $30.7\%$ of the overall sample), and subsequently adding observations for persons of other races that we subsampled with probabilities proportional to their NHANES weights. This resulted in a sample ($n = 5,\!058$) composed of about $60.6\%$ of White persons and $49.3\%$ females. We restricted the dataset to adults aged 18-65 years and removed 378 incomplete observations. The final sample size for analysis was $n = 2,\!146$. Figure~\ref{fig:5} shows the marginal distribution and mid-quantile function of number of prescription medicines, while variables used for analysis are summarized in Table~\ref{tab:9}.

\begin{figure}
\centering
\includegraphics[scale = 0.55]{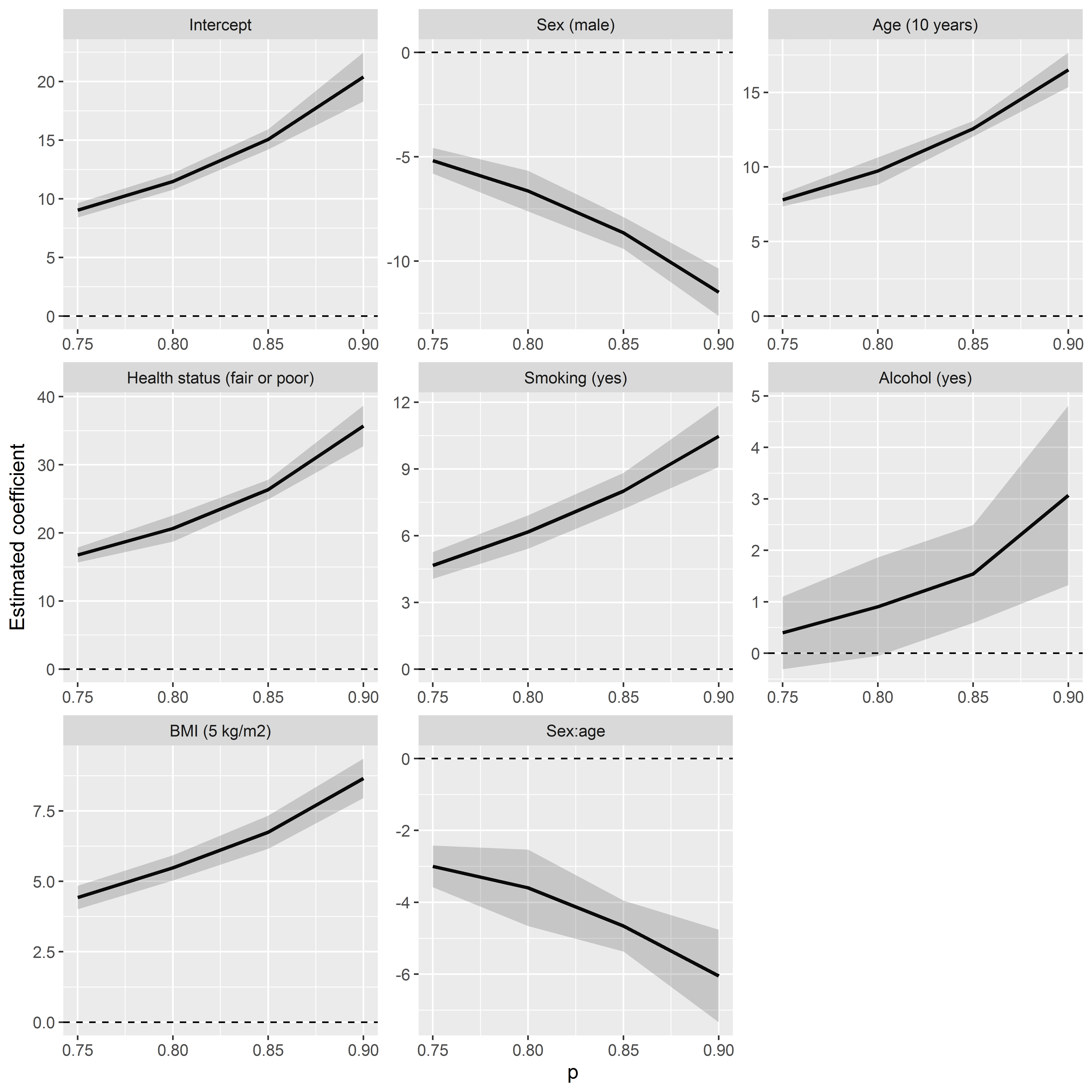}
\caption{Estimated mid-quantile regression coefficients ($p \in \{0.75, 0.8, 0.85, 0.9\}$) and point-wise $95\%$ confidence bands for the number of prescription medicines in the United States using the National Health and Nutrition Examination Survey data, 2015-2016. \label{fig:6}}
\end{figure}

We investigated a linear model with sex (baseline: female), age (centered at 40 and scaled by 10), health status (baseline: good or excellent), smoking status (baseline: no), alcohol use (baseline: no), body mass index---BMI (centered at 30 and scaled by 5), and the interaction between sex and age. The admissible range $\mathcal{I}$ for the application of \eqref{eqn:9} was $[0.46, 0.92]$, thus resulting is a large value of the lower bound due to the high proportion of zeros in the response. In Figure~\ref{fig:6}, we report the estimates of the coefficients obtained via \eqref{eqn:9} and their $95\%$ confidence intervals based on \eqref{eqn:10} for $p \in \{0.75, 0.8, 0.85, 0.9\}$. Older males tend to use less medications as compared to their female peers, and the inequality is more marked among high users (i.e., larger values of $p$), with the largest difference of 27 prescriptions between two 65-year-olds of opposite sex ranking in the top $10\%$ of their respective distributions. However, due to a negative sex-age interaction, the inequality is actually reversed at younger ages, though differences are very small. Put differently, prescription medications count increases with age (as expected) and the rate of increase is sex-specific (steeper for females) and quantile-dependent (steeper at higher $p$). A fair or poor perceived health status is associated with an estimated 17 ($p = 0.75$) to 36 ($p = 0.9$) more prescriptions than what is expected with a good or excellent health status. Individuals classified as smokers have an estimated 5 ($p = 0.75$) to 10 ($p = 0.9$) more prescriptions than non-smokers. However, one should be cautious with the interpretation of this result since the smoking variable used here does not capture smoking habits, past or recent. Alcohol, too, increases the prescription medications count and is quantile-dependent, though the magnitudes of the estimated effects are smaller than those associated with other factors. Finally, higher BMI is associated with a considerable larger number of prescription medications. For every 5 kg/m$^2$ increase in BMI, it is estimated that the 75th mid-quantile increases by about 4 prescriptions, while the 90th mid-quantile increases by about 9 prescriptions.

\begin{landscape}
\centering
\begin{figure}
\includegraphics[scale = 0.6]{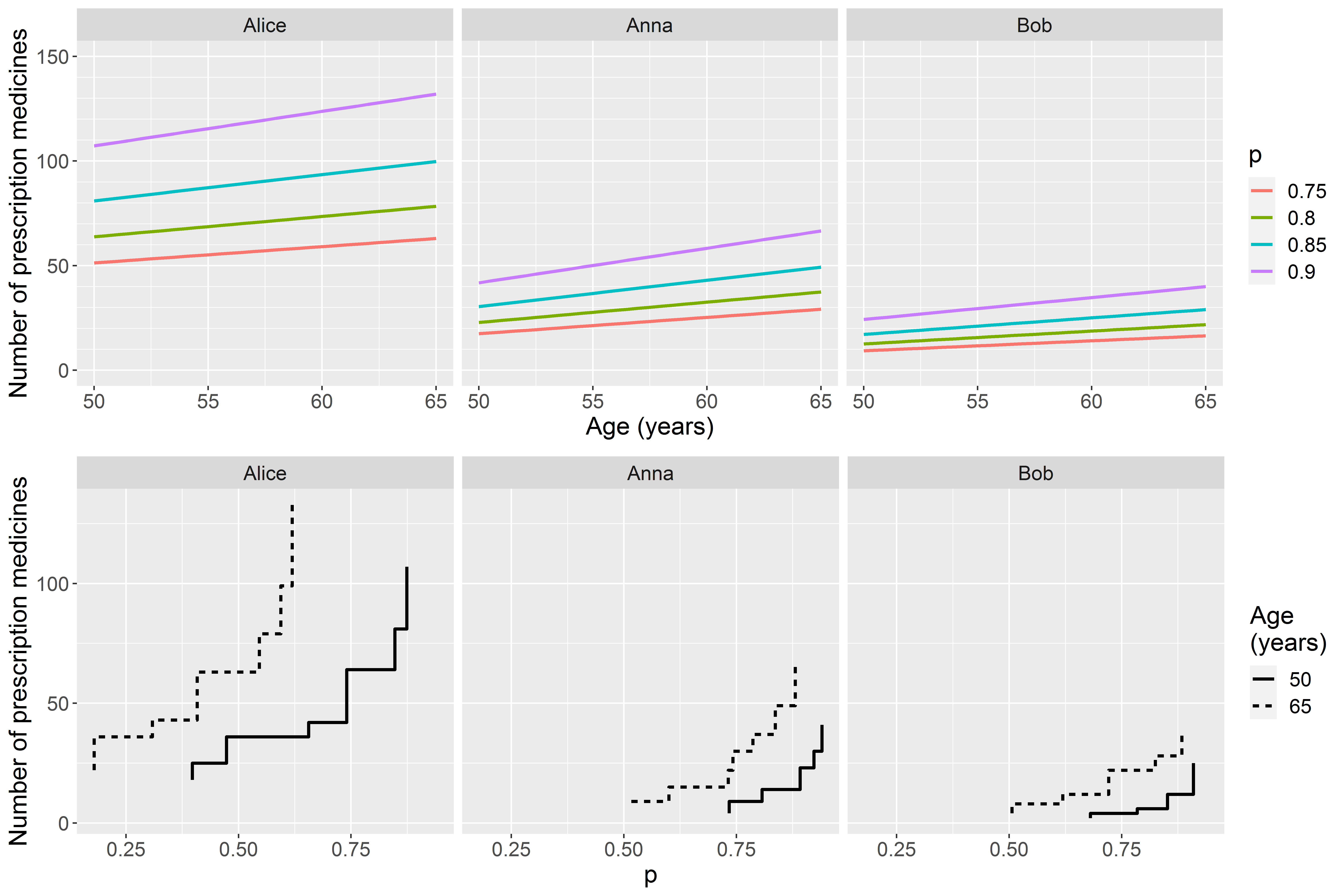}
\caption{Estimated conditional mid-quantiles (top row) and ordinary quantile functions (bottom row) of the number of prescription medicines for three hypothetical subpopulation profiles using the National Health and Nutrition Examination Survey data, 2015-2016. \label{fig:7}}
\end{figure}
\end{landscape}

To exemplify the results, we defined three hypothetical subpopulation profiles: Alice, a woman with fair or poor health status, non-smoker, non-alcohol-user, with a BMI equal to 50; Anna, a woman with good or excellent health status, smoker, alcohol user, with a BMI equal to 25; and Bob, a man with same attributes as Anna's. In Figure~\ref{fig:7}, we show the estimated mid-quantiles as a function of age (50--65 years) for the three subpopulations. Our model estimates that women like Alice who are in the top $25\%$ of their subpopulation's distribution use as much as 51 prescription medications by the age of 50. The count raises to about 107 for those who are in the top $10\%$, with an additional 25 prescriptions when reaching the age of 65. Women like Anna, who enjoy a better health status and a lower BMI, are relatively better off than women like Alice, with substantially lower prescription drugs at all quantiles and all ages. Finally, as we noted above, estimates for (older) men start from lower values and increase more slowly with increasing age as compared to women. More specifically, a 65-year-old Bob in the top $10\%$ of his subpopulation's distribution has approximately the same number of prescriptions as a \textit{50 year-old} Anna who ranks similarly in her subpopulation's distribution, all else being equal. In Figure~\ref{fig:7}, we also show the estimated ordinary quantile functions at age 50 and 65 that were recovered from the estimated conditional mid-quantiles following the procedure discussed in Section~\ref{sec:2.4}.

\section{Discussion}
\label{sec:4}
We developed an approach to conditional quantile estimation with discrete responses. We established the theoretical properties of our conditional mid-quantile estimator under general conditions and showed its good performance in a simulation study with data generated from different discrete response models. Our two-step estimator is easy to implement. When constraining the quantile index to a data-driven admissible range, the second-step estimating equation has a least-squares type, closed-form solution, which is computationally efficient.

We note that the nonstandard rate of convergence given in Theorem~\ref{theo:2} may dampen the enthusiasm for the closed-form solution of estimator \eqref{eqn:9}. However, faster rates of alternative estimators may come at a price. The $\sqrt{n}$-convergence of \citeauthor{Machado2005}'s (\citeyear{Machado2005}) estimator, for example, is restricted to finding an $n$-sequence of real numbers that depends on the model's transformation to linearity (see their assumption A6). This means that implementation of jittering requires ad hoc asymptotic calculations and software programming, neither of which are typically source of excitement for the applied users. In general, while our proposed estimator does not achieve the much-coveted convergence rate of classical estimators, we can take solace in the fact that its bias and loss of efficiency can be ignored already at moderate sample sizes. As compared to jittering-based quantile regression, mid-quantile regression allows for the modeling of a wider range of discrete outcomes and it shows efficiency gains already at small sample sizes. Last but not least, the availability of mid-quantile regression in the \texttt{R} package \texttt{Qtools} \citep{Geraci2016}, of which a brief tutorial is given in Appendix~\ref{sec:B}, may be particularly attractive in applied research studies.

In a real data analysis, conditional mid-quantiles revealed interesting aspects of prescription drugs use in the US. In general, our results on gender and age inequalities are consistent with the literature \citep{Roe2002,Loikas2013}. While differences in medications use between men and women are expected due to differences in incidence of disease (e.g., genitourinary infections, migraines, impaired thyroid function) or biological differences or differential preventive healthcare use, gender inequalities may be due also to unequal treatment \citep{Loikas2013}. Our analysis provides additional insight as it shows that differences between men and women (especially among older individuals) are heterogeneous, with a substantially higher gender gap among the top $10\%$ of the distribution, after adjusting for perceived health status and other covariates. This raises the alarm about a possible differential medical treatment that worsens the inequality among the most fragile segment of the population. On the other hand, the heterogeneous association between number of prescription medicines and BMI may be easier to explain on medical grounds. Higher prescribing levels are expected to be associated with higher BMI because of medical conditions that are known to be more prevalent in obese individuals \citep{CPT2005}. If a higher percentile of the medications count distribution is interpreted as an (objective) index of poor health, then it is natural to expect that a given increase in BMI is more detrimental for someone ranking at, say, the 90th percentile than it is for the `average' prescription drugs user. Owing to the fact that obesity increases drug prescribing in the most expensive prescribing categories \citep{CPT2005}, it becomes clear that intervening on this modifiable risk factor has important connotations also for controlling healthcare expenditure.

We believe that mid-quantile regression is amenable to several possible extensions, including estimation in the presence of censoring (survival analysis) and clustering (e.g., longitudinal analysis). Further research is needed to develop computationally efficient methods for high-dimensional data. Useful hints on how to tackle issues regarding censoring and higher-dimensional covariates may be gathered from the work by \cite{deBacker}.

\section*{Acknowledgements}
Mid-quantile regression is implemented in the \texttt{R} \citep{R} package \texttt{Qtools} \citep{Geraci2016}. The latter also includes routines to fit quantile regression for counts as proposed by \cite{Machado2005}.

\clearpage

\appendix

\section{Supplementary theoretical results}
\label{sec:A}
\numberwithin{equation}{section}
\setcounter{equation}{0}

In this section, we prove Theorems~\ref{theo:1} and~\ref{theo:2}. We begin by providing some auxiliary results. We assume, throughout, that $\hat{G}_{Y|X}^{c}(\cdot|x)$ is a linear interpolant. While the validity of the Theorems still holds for other types of interpolants (e.g., polynomial), analytical expressions are more tractable in the linear case.

\subsection{Auxiliary results}
\label{sec:A.1}
Our objective function and estimator are given by
\begin{equation}\label{eqn:A1}
\psi_{n}(\beta; p) = \frac{1}{n}\sum_{i = 1}^{n} \left\{p - \hat{G}_{Y|X}^{c}\left(\eta_{i}|x_{i}\right)\right\}^{2}
\end{equation}
and
\begin{equation}\label{eqn:A2}
\hat{\beta}(p) = \argmin_{\beta \in \mathbb{R}^{q}} \psi_{n}(\beta; p),
\end{equation}
respectively. The equation of the interpolating function can be written explicitly as
\begin{eqnarray*}
\hat{G}_{Y|X}^{c}\left(\eta_{i}|x_{i}\right) = b_{j_{i}}(\eta_{i} - z_{j_{i}}) + \hat{\pi}_{j_{i}} & & \qquad \text{$z_{j_{i}} \leq \eta_{i} \leq z_{j_{i}+1}$,}
\end{eqnarray*}
where $b_{j_{i}} = \dfrac{\hat{\pi}_{j_{i}+1} - \hat{\pi}_{j_{i}}}{z_{j_{i}+1} - z_{j_{i}}}$ and $\hat{\pi}_{j_{i}} = \hat{G}_{Y|x_{i}}(z_{j_{i}})$. The index $j_{i} = 1,\ldots,k-1$ identifies, for a given  $i = 1, \ldots, n$, the value $z_{j_{i}}$ among the $z$'s such that $\hat{G}_{Y|x_{i}}(z_{j_{i}}) \leq p \leq \hat{G}_{Y|x_{i}}(z_{j_{i}+1})$.

Then, the derivative of $\psi_{n}$ with respect to the $h$th element of $\beta$ is given by
\begin{equation*}
\frac{\partial\psi_{n}(\beta; p)}{\partial\beta_h} = \frac{1}{n}\sum_{i = 1}^{n} 2\left\{p - \hat{G}_{Y|X}^{c}\left(h^{-1}(x_i\tp\beta)|x_i\right)\right\}\left\{- \frac{\partial \hat{G}_{Y|X}^{c}\left(h^{-1}(x_i\tp\beta)|x_i\right)}{\partial\beta_h}\right\},
\end{equation*}
where
\begin{equation*}
\frac{\partial \hat{G}_{Y|X}^{c}\left(h^{-1}(x_i\tp\beta)|x_i\right)}{\partial\beta_h} = x_{ih}b_{j_{i}}\frac{\partial h^{-1}(\eta_{i})}{\partial\eta_{i}},
\end{equation*}
the existence of which follows from the differentiability of $h$.

Now, consider the second derivative of the objective function
\begin{align*}
  \frac{\partial^2\psi_{n}(\beta; p)}{\partial\beta_h\partial\beta_u} = & - \frac{2}{n}\sum_{i = 1}^{n} \left[p - \hat{G}_{Y|X}^{c}\left\{h^{-1}(x_i\tp\beta)|x_{i}\right\}\right] \frac{\partial^2 \hat{G}_{Y|X}^{c}\left\{h^{-1}(x_i\tp\beta)|x_{i}\right\}}{\partial\beta_h\partial\beta_u}\\
  & - \frac{\partial \hat{G}_{Y|X}^{c}\left\{h^{-1}(x_i\tp\beta)|x_{i}\right\}}{\partial\beta_h}\frac{\partial \hat{G}_{Y|X}^{c}\left\{h^{-1}(x_i\tp\beta)|x_{i}\right\}}{\partial\beta_u},
\end{align*}
where
\begin{equation*}
\frac{\partial^2 \hat{G}_{Y|X}^{c}\left(h^{-1}(x_i\tp\beta)|x_{i}\right)}{\partial\beta_h\partial\beta_u} = x_{ih}x_{iu}b_{j_{i}}\frac{\partial^2 h^{-1}(\eta_{i})}{\partial\eta_{i}}.
\end{equation*}
In summary, we obtain
\begin{equation*}
  \frac{\partial^2\psi_{n}(\beta; p)}{\partial\beta_h\partial\beta_u} = - \frac{2}{n}\sum_{i = 1}^{n} x_{ih}x_{iu}b_{j_{i}}\left[p - \hat{G}_{Y|X}^{c}\left\{h^{-1}(x_i\tp\beta)|x_{i}\right\}\right] \frac{\partial^2 h^{-1}(\eta_{i})}{\partial\eta_{i}} - x_{ih}x_{iu}\left\{b_{j_{i}}\frac{\partial h^{-1}(\eta_{i})}{\partial\eta_{i}}\right\}^2.
\end{equation*}
Clearly, if $h$ is the identity function, then
\begin{equation*}
  \frac{\partial^2\psi_{n}(\beta; p)}{\partial\beta_h\partial\beta_u} = \frac{2}{n}\sum_{i = 1}^{n} x_{ih}x_{iu}b_{j_{i}}^2.
\end{equation*}

\subsection{Proof of Theorem~\ref{theo:1}}
\label{sec:A.2}
\begin{proof}
Under the conditions stated \citep{Li2008}
\begin{equation*}
\max_z \left|\hat G_{Y|X}(z) - G_{Y|X}(z)\right| \to 0
\end{equation*}
as $n \to \infty$. We can also verify that
\begin{equation*}
\sup_z \left|\hat G^{c}_{Y|X}(z|x) - G^{c}_{Y|X}(z|x)\right| \to 0.
\end{equation*}
Consequently,
\begin{equation*}
\Pr\left(\lim_n \hat G^{c}_{Y|X}\left[h^{-1}\left\{x\tp\beta(p)\right\}|x\right] = G^{c}_{Y|X}\left[h^{-1}\left\{x\tp\beta(p)\right\}|x\right]\right) = 1.
\end{equation*}
Consider now $\gamma(p) \neq \beta^{*}(p)$. It is straightforward to verify that
\begin{equation*}
\left(p-G^{c}_{Y|X}\left[h^{-1}\left\{x\tp\beta^{*}(p)\right\}|x\right]\right)^2 \leq \left(p-G^{c}_{Y|X}\left[h^{-1}\{x\tp\gamma(p)\}|x\right]\right)^2.
\end{equation*}
In fact, if $h^{-1}\left\{x\tp\beta^{*}(p)\right\}=y_j$ for some value of $p$ and $y_{j} \in \mathcal{S}_{Y}$, then $(p-G^{c}_{Y|X}[h^{-1}\{x\tp\beta^{*}(p)\}|x])^2=0$; while all other values are obtained through interpolation. A consequence is that $\beta^{*}(p)$ is, eventually, a solution of the minimization problem in \eqref{eqn:2}. Additionally, there is only one such solution, since, by assumption, $\Pr(Y = y|X) > 0$ for all $y \in \mathcal{S}_{Y}$, and $G^{c}(\eta(p)|x)$ is monotonic for $\pi_{1} < p < \pi_{k}$, where $\pi_{1}$ and $\pi_{k}$ are the mid-probabilities corresponding to, respectively, the smallest and largest discrete value (if $k = \infty$, then $\pi_{1} < p < 1$). This implies consistency of $\hat\beta(p)$, the minimizer in \eqref{eqn:2}. Consistency of the predicted mid-quantiles follows directly.
\end{proof}

\subsection{Proof of Theorem~\ref{theo:2}}
\label{sec:A.3}
\begin{proof}
Since the differentiability of $\psi_n(\beta;p)$ follows from the assumptions, we can apply a first-order Taylor expansion to obtain
\begin{equation}\label{eqn:A3}
\nabla_{\beta} \psi_n(\hat\beta;p) = \nabla_{\beta} \psi_n(\beta^{*};p) +
\nabla^2_{\beta} \psi_n(\beta^{+};p)(\hat\beta-\beta^{*}),
\end{equation}
where $\beta^{+}$ is a point in the interior of the hypercube delimited by $\hat\beta$ and $\beta^{*}$. Expressions for $\nabla_{\beta} \psi_n$ and $\nabla^2_{\beta} \psi_n$ are given in Section~\ref{sec:A.1}. Note that $\nabla_{\beta} \psi_n(\hat\beta;p)=0$ since $\hat\beta$ is the minimizer in \eqref{eqn:A2}. The assumption on the design matrix guarantees that the Hessian $\nabla^{2}_{\beta} \psi_n(\beta^{+};p)$ is positive definite. Hence, we can rewrite \eqref{eqn:A3} as
\begin{equation}
\label{eqn:A4}
\sqrt{n\prod_j\lambda_j} (\hat\beta-\beta^{*}) = -(\nabla^2_{\beta} \psi_n(\beta^{+};p))^{-1}\sqrt{n\prod_j\lambda_j}  \nabla_{\beta} \psi_n(\beta^{*};p).
\end{equation}

To derive the asymptotic distribution of $\hat\beta$, it suffices to study the asymptotic distribution of the right-hand side of \eqref{eqn:A4}. First, let $J(b)=E\left\{\nabla^2_{\beta} \psi_n(\beta;p)\Big\rvert_{\beta = b}\right\}$. By using the consistency results in Theorem~\ref{theo:1} and the triangle inequality, it is immediate to show that $\nabla^2_{\beta} \psi_n(\beta^{+};p)$ weakly converges element-wise to $J(\beta^{*})$. Using the results in Section~\ref{sec:A.1}, we then can write
\begin{align*}
\sqrt{n\prod_j\lambda_j} \nabla_{\beta} \psi_n(\beta^{*};p)  = & -2\sqrt{\frac{1}{n}\prod_j\lambda_j} \sum_{i=1}^n \nabla_{\beta} \hat G^{c}_{Y|X}\left\{h^{-1}(x_i\tp\beta^{*})|x\right\} \\
& \times \left[p-\hat G^{c}_{Y|X}\left\{h^{-1}(x_i\tp\beta^{*})|x\right\}\right].
\end{align*}
We need to demonstrate that the expression above converges in distribution, thus we expand the quantities on the right-hand side as follows:
\begin{align}
  \label{eqn:A5}
  \nonumber \sqrt{n\prod_j\lambda_j} \nabla_{\beta} \psi_n(\beta^{*};p)  =& -\frac{2}{n} \sum_{i=1}^n x_i \dot{h}^{-1}(\eta_i) \frac{p}{z_{j_i+1}-z_{j_i}} \sqrt{n\prod_j\lambda_j} \hat G_{Y|x_i}(z_{j_i+1})\\
  \nonumber &+ \frac{2}{n} \sum_{i=1}^n x_i \dot{h}^{-1}(\eta_i) \frac{p}{z_{j_i+1}-z_{j_i}} \sqrt{n\prod_j\lambda_j} \hat G_{Y|x_i}(z_{j_i}) \\
  \nonumber &+ \frac{2}{n} \sum_{i=1}^n x_i \dot{h}^{-1}(\eta_i) \frac{\hat{G}^{c}_{Y|X}\left\{h^{-1}(x_i\tp\beta^{*})|x_{i}\right\}}{z_{j_i+1}-z_{j_i}}  \sqrt{n\prod_j\lambda_j} \hat G_{Y|x_i}(z_{j_i+1})\\
  &- \frac{2}{n} \sum_{i=1}^n x_i \dot{h}^{-1}(\eta_i) \frac{\hat{G}^{c}_{Y|X}\left\{h^{-1}(x_i\tp\beta^{*})|x_{i}\right\}}{z_{j_i+1}-z_{j_i}}  \sqrt{n\prod_j\lambda_j} \hat G_{Y|x_i}(z_{j_i}),
\end{align}
where $\dot{h}^{-1}(\eta_i) = \frac{\partial h^{-1}(\eta_i)}{\partial \eta_i}$. First of all, as shown in \cite{Li2008}, $\sqrt{n\prod_j\lambda_j} \hat G_{Y|x_i}(z_{j_i})$ converges in distribution to a Gaussian random variable for all $i$. Additionally, the assumptions on the bandwidths guarantee asymptotic independence of $\hat G_{Y|x_h}(z)$ and $\hat G_{Y|x_l}(z)$ for $x_l \neq x_h$ and all $z$. To see this, note that $K_{\lambda}(X_i,x) \to 0$ for all $X_i \neq x$. According to the dominated convergence theorem, the asymptotic covariance of $\hat G_{Y|x_h}(z)$ and $\hat G_{Y|x_l}(z)$ is zero. Asymptotic independence follows by the Cramer-Wold device. Furthermore, $\Pr(z_{j_i+1}-z_{j_i} \neq 0)=1$ since $Y$ is discrete. Finally, note that by our Theorem~\ref{theo:1}$, \hat G^{c}_{Y|X}\left\{h^{-1}(x_i\tp\beta^{*})|x_{i}\right\}$ converges in probability to a constant value. By combining the results above with the assumptions on the design matrix (namely, that $1/n \sum_i x_i$ converges to a bounded vector), we obtain convergence in distribution of the right-hand side of \eqref{eqn:A5} to a Gaussian random variable.

Therefore, $\sqrt{n\prod_j\lambda_j} \nabla_{\beta} \psi_n(\beta^{*};p)$ is asymptotically normal with variance
\begin{equation}
\label{eqn:A6}
D(\beta^{*}) = {\rm Var}\left( \frac{2\sqrt{\prod_j \lambda_j}}{\sqrt{n}} \sum_{i=1}^n \nabla_{\beta} \hat G^{c}_{Y|X}\left\{h^{-1}(x_i\tp\beta^{*})|x_{i}\right\} \left[p-\hat G^{c}_{Y|X}\left\{h^{-1}\left(x_i\tp\beta^{*}\right)|x_{i}\right\}\right]\right).
\end{equation}
By letting
\begin{equation}
\label{eqn:A7}
V(\beta^{*}) = J(\beta^{*})^{-1}D(\beta^{*})J(\beta^{*})^{-1},
\end{equation}
we obtain
\begin{equation*}
V(\beta^{*})^{-1/2} \sqrt{n}(\hat\beta-\beta^{*}) \stackrel{d}{\to} N(0,I_q).
\end{equation*}
\end{proof}

A consistent estimator of $V(\beta^{*})$ could be found by calculating sample averages of the quantities involved in $J(\beta^{*})$, and computing $D(\beta^{*})$ via resampling. However, using expression~\eqref{eqn:10} leads to an analytical calculation of the variance of $\hat{\beta}$ with clear computational advantages.

\clearpage

\section{Supplementary simulation study results}
\label{sec:B}
\numberwithin{equation}{section}
\setcounter{equation}{0}
\renewcommand{\thetable}{B\arabic{table}}
\setcounter{table}{0}

\begin{table}[ht]
\caption{Bias and root mean squared error (RMSE) of predicted quantiles for data generated using the homoscedastic discrete uniform model (1b). \label{tab:B1}}
\begin{tabular}{lrrrrrrr}
\hline
 & \multicolumn{2}{c}{$n = 100$} & \multicolumn{2}{c}{$n = 500$} & \multicolumn{2}{c}{$n = 1000$} & \\
$p$ & \multicolumn{1}{c}{Bias} & \multicolumn{1}{c}{RMSE} & \multicolumn{1}{c}{Bias} & \multicolumn{1}{c}{RMSE} & \multicolumn{1}{c}{Bias} & \multicolumn{1}{c}{RMSE} & \multicolumn{1}{c}{$\bar{H}$}\\
\hline
0.2 & $-$0.046 & 0.803 & $-$0.037 & 0.528 & $-$0.036 & 0.453 & 8.995 \\
  0.3 & 0.071 & 0.827 & 0.016 & 0.535 & 0.000 & 0.456 & 9.995 \\
  0.4 & 0.122 & 0.849 & 0.034 & 0.537 & 0.014 & 0.455 & 10.995 \\
  0.5 & 0.156 & 0.854 & 0.046 & 0.532 & 0.022 & 0.451 & 11.995 \\
  0.6 & 0.197 & 0.851 & 0.055 & 0.521 & 0.031 & 0.439 & 12.995 \\
  0.7 & 0.245 & 0.837 & 0.067 & 0.507 & 0.041 & 0.425 & 13.995 \\
  0.8 & 0.346 & 0.839 & 0.111 & 0.491 & 0.069 & 0.412 & 14.995\\
\hline
\end{tabular}
\end{table}

\begin{table}[ht]
\caption{Bias and root mean squared error (RMSE) of predicted quantiles for data generated using the heteroscedastic discrete uniform model (2b). \label{tab:B2}}
\begin{tabular}{lrrrrrrr}
\hline
 & \multicolumn{2}{c}{$n = 100$} & \multicolumn{2}{c}{$n = 500$} & \multicolumn{2}{c}{$n = 1000$} & \\
$p$ & \multicolumn{1}{c}{Bias} & \multicolumn{1}{c}{RMSE} & \multicolumn{1}{c}{Bias} & \multicolumn{1}{c}{RMSE} & \multicolumn{1}{c}{Bias} & \multicolumn{1}{c}{RMSE} & \multicolumn{1}{c}{$\bar{H}$}\\
\hline
0.2 & $-$0.463 & 1.838 & $-$0.324 & 1.227 & $-$0.344 & 1.114 & 13.988 \\
  0.3 & $-$0.545 & 2.167 & $-$0.394 & 1.462 & $-$0.390 & 1.343 & 16.986 \\
  0.4 & $-$0.562 & 2.431 & $-$0.457 & 1.719 & $-$0.431 & 1.591 & 19.983 \\
  0.5 & $-$0.501 & 2.662 & $-$0.507 & 1.972 & $-$0.463 & 1.848 & 22.981 \\
  0.6 & $-$0.228 & 2.857 & $-$0.474 & 2.211 & $-$0.461 & 2.104 & 25.978 \\
  0.7 & 0.175 & 3.060 & $-$0.275 & 2.455 & $-$0.300 & 2.353 & 28.976 \\
  0.8 & 0.843 & 3.376 & 0.196 & 2.749 & 0.108 & 2.659 & 31.973\\
\hline
\end{tabular}
\end{table}

\begin{table}[ht]
\caption{Bias and root mean squared error (RMSE) of predicted quantiles for data generated using the Poisson model (3b). \label{tab:B3}}
\begin{tabular}{lrrrrrrr}
\hline
 & \multicolumn{2}{c}{$n = 100$} & \multicolumn{2}{c}{$n = 500$} & \multicolumn{2}{c}{$n = 1000$} & \\
$p$ & \multicolumn{1}{c}{Bias} & \multicolumn{1}{c}{RMSE} & \multicolumn{1}{c}{Bias} & \multicolumn{1}{c}{RMSE} & \multicolumn{1}{c}{Bias} & \multicolumn{1}{c}{RMSE} & \multicolumn{1}{c}{$\bar{H}$}\\
\hline
0.2 & $-$18.167 & 36.048 & $-$13.598 & 27.366 & $-$12.075 & 24.612 & 216.351 \\
  0.3 & $-$9.786 & 23.088 & $-$8.372 & 19.552 & $-$7.651 & 17.930 & 220.421 \\
  0.4 & $-$3.072 & 14.097 & $-$4.141 & 13.343 & $-$3.996 & 12.624 & 223.926 \\
  0.5 & 3.416 & 11.066 & 0.076 & 7.978 & $-$0.371 & 7.560 & 227.223 \\
  0.6 & 9.946 & 14.939 & 4.761 & 7.626 & 3.551 & 6.309 & 230.542 \\
  0.7 & 16.268 & 21.987 & 9.473 & 12.590 & 7.860 & 10.549 & 234.117 \\
  0.8 & 26.405 & 35.420 & 15.210 & 20.174 & 12.691 & 16.860 & 238.331\\
\hline
\end{tabular}
\end{table}

\begin{table}[ht]
\caption{Bias and root mean squared error (RMSE) of predicted quantiles for data generated using the Bernoulli model (4b). \label{tab:B4}}
\begin{tabular}{lrrrrrrr}
\hline
 & \multicolumn{2}{c}{$n = 100$} & \multicolumn{2}{c}{$n = 500$} & \multicolumn{2}{c}{$n = 1000$} & \\
$p$ & \multicolumn{1}{c}{Bias} & \multicolumn{1}{c}{RMSE} & \multicolumn{1}{c}{Bias} & \multicolumn{1}{c}{RMSE} & \multicolumn{1}{c}{Bias} & \multicolumn{1}{c}{RMSE} & \multicolumn{1}{c}{$\bar{H}$}\\
\hline
0.5 & $-$0.000 & 0.067 & 0.000 & 0.029 & 0.000 & 0.021 & 0.577 \\
   \hline
\end{tabular}
\end{table}

\clearpage

\section{R code}
\label{sec:C}
\numberwithin{equation}{section}
\setcounter{equation}{0}

In this section, we provide an example on how to do inference on mid-quantile regression models using the R package \texttt{Qtools} \citep{Geraci2016}. The latter is available on CRAN and can be installed as follows:
\begin{verbatim}
install.packages("Qtools")
\end{verbatim}

\noindent We consider the dataset \texttt{esterase}, which is available in the Qtools package. The dataset contains data from an essay for the concentration of an enzyme esterase. The observed concentration of esterase was recorded (\texttt{esterase}), and then in a binding experiment the number of bindings were counted (\texttt{Count}). After loading the package, the following code shows how to attach the dataset and access the R documentation describing the variables:
\begin{verbatim}
library(Qtools)
data(esterase)
?esterase

> head(esterase)
  Esterase Count
1      3.1    28
2      5.6   166
3      6.1    52
4      6.4    84
5      6.5    85
6      6.7    86
\end{verbatim}

\noindent We estimate the marginal mid-quantiles of the discrete variable \texttt{Count} using the function \texttt{midquantile}.
\begin{verbatim}
fit <- midquantile(esterase$Count, probs = 1:3/4)

> str(fit)
List of 5
 $ call: language midquantile(x = esterase$Count, probs = 1:3/4)
 $ x   : num [1:3] 0.25 0.5 0.75
 $ y   : num [1:3] 147 269 419
 $ fn  :function (v)
 $ data: int [1:113] 28 166 52 84 85 86 127 104 107 96 ...
 - attr(*, "class")= chr "midquantile"
\end{verbatim}
The output is a list that contains the estimated mid-quantiles (\texttt{y}) at the specified probabilities (\texttt{x}). It also contains the interpolating mid-quantile function (\texttt{fn}) which can be plotted using the associated \texttt{plot.midquantile} function. Confidence intervals for mid-quantile estimates can be obtained using \texttt{confint.midquantile}.

\noindent Suppose we want to fit the linear model $H(p) = \beta_{0} + \beta_{1}(p)x$ to estimate the 0.25 and 0.75 conditional mid-quantiles of \texttt{Count} as a function of \texttt{esterase}. We use the main command \texttt{midrq} where the argument \texttt{tau} specifies the level of the quantiles of interest.
\begin{verbatim}
fit <- midrq(Count ~ Esterase, tau = c(0.25, 0.75), data = esterase,
type = 3, control = midrqControl(method = "Nelder-Mead", ecdf_est = "npc"))

> fit
call:
midrq(formula = Count ~ Esterase, data = esterase, tau = c(0.25,
    0.75), type = 3, control = midrqControl(method = "Nelder-Mead",
    ecdf_est = "npc"))


Coefficients linear predictor:
                 0.25     0.75
(Intercept) -48.97063 16.02915
Esterase     15.61743 19.12168

Degrees of freedom: 113 total; 111 residual
\end{verbatim}

\noindent There are three estimators available in \texttt{midrq} and these can be selected via the argument \texttt{type}. Using \texttt{type = 1}, the minimization of the objective function \eqref{eqn:7} is carried out using a general purpose optimizer (by default, this is Nelder-Mead, although it can be changed via \texttt{midrqControl}). When \texttt{type = 2}, optimization is based on a CUSUM process (which is not discussed in the present work and should be considered experimental). Finally, \texttt{type = 3} gives the least-squares-type estimator in equation \eqref{eqn:9}. On the other hand, the argument \texttt{ecdf\_est} in \texttt{midrqControl} controls the conditional mid-CDF estimator (for example, \texttt{ecdf\_est = "npc"} gives the kernel estimator by \cite{Hayfield2008}).

\noindent The package provides several S3 methods for fitted \texttt{midrq} objects including: \texttt{summary}, which gives standard errors, $p$-values, and confidence intervals; \texttt{coef} to extract estimates of the regression coefficients; \texttt{vcov} to extract the variance-covariance matrix of the estimator $\hat{\beta}(p)$ defined in Section~\ref{sec:2.3}; and \texttt{predict} and \texttt{residuals}, whose names are self-explanatory. The function \texttt{midq2q} gives an estimate of ordinary quantiles using the procedure described in Section~\ref{sec:2.4}. Finally, we draw attention on the availability in the \texttt{Qtools} package of the functions \texttt{midecdf} and \texttt{cmidecdf} for estimating marginal and conditional mid-cumulative probabilities, respectively.


\end{document}